\pdfoutput=1 % ensure pdflatex
\documentclass[11pt, twoside]{mathdoc}
\usepackage{distoperators}

\usepackage[title]{appendix}
\usepackage{layouts}
\usepackage{IEEEtrantools}
\usepackage[shortlabels, inline]{enumitem}
\usepackage{bm}
\usepackage{mathrsfs}
\usepackage{listings}

\usepackage{booktabs}
\usepackage{multirow}
\usepackage{tikz}
\usetikzlibrary{bayesnet}
\usetikzlibrary{arrows}
\usepackage{placeins}
\usepackage{multirow}
\usepackage{microtype}
\usepackage{amsthm}
\usepackage[frozencache,cachedir=minted-cache]{minted}
\usepackage[T1]{fontenc}

\setminted{
    frame=lines,
    framesep=2mm,
    bgcolor=white,
    fontsize=\tiny,
    linenos,
    breaklines
}

\definecolor{arr2flatcolour}{HTML}{4477AA}
\definecolor{arr2minncolour}{HTML}{EE6677}\definecolor{arr2bumpcolour}{HTML}{66CCEE}
\definecolor{gaussiancolour}{HTML}{228833}
\definecolor{minnesotacolour}{HTML}{CCBB44}
\definecolor{rhscolour}{HTML}{AA3377}

\theoremstyle{plain}

\newtheorem{theorem}{Theorem}[section]

\theoremstyle{definition}
\newtheorem{definition}[theorem]{Definition}

\theoremstyle{remark}

\DeclareMathOperator{\ar}{AR}
\DeclareMathOperator{\arx}{ARX}

\DeclareMathOperator{\ma}{MA}

\DeclareMathOperator{\cpi}{CPI}
\DeclareMathOperator{\arma}{ARMA}
\DeclareMathOperator{\ardl}{ARDL}
\DeclareMathOperator{\VAR}{VAR}
\newcommand{\betaprime}{\beta^\prime}

\newcommand{\tr}[1]{\text{\normalfont tr}\left(#1\right)}

\let\proglang=\textsf
\title{The ARR2 prior: flexible predictive prior definition for Bayesian auto-regressions}
\author[1]{David Kohns}
\author[1]{Noa Kallioinen}
\author[1,2]{Yann McLatchie}
\author[1]{Aki Vehtari}
\affil[1]{Department of Computer Science, Aalto University}
\affil[2]{Department of Statistical Science, University College London}

\makeatletter
\def\runauthor{Kohns, Kallioinen, McLatchie, and Vehtari}
\let\inserttitle\@title
\def\runtitle{The ARR2 prior}
\makeatother

\begin{document}
\maketitle
\thispagestyle{empty}
\begin{abstract}
    We present the ARR2 prior, a joint prior over the auto-regressive components in Bayesian time-series models and their induced $R^2$. 
Compared to other priors designed for times-series models, the ARR2 prior allows for flexible and intuitive shrinkage.
We derive the prior for pure auto-regressive models, and extend it to auto-regressive models with exogenous covariates, and state-space models. 
Through both simulations and real-world modelling exercises, we demonstrate the efficacy of the ARR2 prior in improving sparse and reliable inference, while showing greater inference quality and predictive performance than other shrinkage priors. 
An open-source implementation of the prior is provided.
\end{abstract}
\section{Introduction}
Independent priors over model components have the undesirable tendency to inflate explained variance with increasing model complexity.
As a remedy to this, the recent literature has advocated for the use of joint priors over model components.
Most notably, the R2D2 prior~\citep{zhang_bayesian_2022} allows the modeller to encode some prior belief on the model's performance as defined by the coefficient of determination, $R^2$, which then trickles down to the parameter level through a mapping between the two.
This idea has since been extended to multi-level models~\citep{aguilar_intuitive_2023}, generalised linear models~\citep[GLMs;][]{yanchenko_r2d2_2021} and spatial models~\citep{yanchenko_r2d2_2023}.
While these priors have found rich application, it is not clear whether and how they can be applied to time-series settings. 
The auto-covariance structures defined by lagged observations and latent dynamics create dependencies between the model parameters that govern auto-correlations and the $R^2$. This necessitates adaptation of the previously presented $R^2$ prior frameworks.
The kernel of this work is an extension of the joint shrinkage framework over $R^2$ to the auto-regressive (AR) components of Bayesian time-series models, including latent state-space models.
Concretely, in this paper we: 
\begin{enumerate}[nosep]
    \item derive the ARR2 prior, a predictively-motivated joint shrinkage prior for AR, AR with exogenous covariates (ARX), and auto-regressive latent state-space models;
    \item discuss how variance decompositions in the ARR2 framework may be set and compare to commonly applied time-series priors;
    \item present an implementation of the ARR2 prior in the probabilistic programming framework \proglang{Stan}~\citep{stan} and,
    \item compare estimation and predictive properties to alternative priors through simulated and real-data experiments.
\end{enumerate}

\subsection{Structure of this paper}
In Section~\ref{sec:relation}, we begin by discussing relevant background and previous work. In Section~\ref{sec:arr2-prior} we motivate and define the $R^2$ for $\ar$ models, and establish connections to the previous literature.
Section~\ref{sec:comp_alt_priors} in particular summarises the properites of the ARR2 prior compared to popular alternative shrinkage priors for $\ar$ models. 
In Section~\ref{sec:arr2-for-arx} we adapt the ARR2 prior to ARX and state-space models. 
We then compare performance in simulated data experiments in Section~\ref{sec:simulations}, and a real-world data forecasting exercise to US inflation in Section~\ref{sec:case-study}. 
Finally, we conclude in Section~\ref{sec:discussion} with some recommendations and discussion of promising future research directions.

\subsection{Background and previous work}
\label{sec:relation}
For a model with many regression coefficients, independent priors, which do not encode dependence between coefficients, can imply a high degree of explained variance~\citep[e.g.,][]{gelmanRegressionOtherStories2020}. In terms of $R^2$, this is exhibited as a prior concentrated near the upper bound of $1$, indicating that the model is expected to almost perfectly fit the data. If the likelihood does not dominate the prior, this can lead to posterior estimates of explained variance that overestimate the actual explanatory power of the model at the population level\footnote{This can be seen by in-sample $R^2$ being much higher than out-of-sample $R^2$.}, negatively impacting both predictive performance and inference.

Imposing sparsity with discrete mixture priors~\citep{ishwaran2005spike} or shrinkage with continuous priors ~\citep{carvalho_handling_2009,bhattacharya2015dirichlet,piironen_sparsity_2017} may limit these tendencies by regularising the variance of the predictor term. 
Notably, the R2D2 prior~\citep{zhang_bayesian_2022} takes the approach of directly encoding prior beliefs about the $R^2$ as a function of total variance and residual variance. 
The prior total variance explained is then decomposed and assigned to the additive components of the model through a simplex prior. By choosing appropriate priors over the simplex, one may also induce sparsity-favouring patterns in coefficients, which is separate from the information encoded by the prior on $R^2$.

A similar prior framework is proposed by \citet{fuglstad_intuitive_2020} in which a variance parameter is shared by all coefficients across model components that is either divided according to a simplex prior or regularised via a penalised complexity prior \citep{simpson2017penalising}. Penalised complexity priors have been extended to AR models by \citet{sorbye_penalised_2017}. This approach specifies priors on partial autocorrelations of the AR process, and does not directly set a prior on $R^2$. Such penalised complexity priors are not straightforwardly extended to the different time-series considered in this paper. As such, we will leave any comparisons to such priors for future investigation.

The previous time-series literature has remained wary of the usefulness of $R^2$ as a measure of model fit. 
This is due to the non-standard frequency distribution of $R^2$ under non-stationarity~\citep{phillips1986understanding}, and the often condemned property of $R^2$, conditional on parameters and design, to increase monotonically with the number of covariates included. 
Here though, we will assume stationarity and are interested in finite sample properties under different priors~\citep[however, see][for an approach to enforce stationarity through the prior]{heaps_enforcing_2022}.
Indeed, we will show that popular priors used in the time-series literature can be extremely informative on the $R^2$ space, despite having no explicit interest in the quantity\footnote{\citet{kovalBayesianReconciliationReturn2024} show however how the Bayesian $R^2$ is a useful metric to describe stock return explainability.}. 
As such, even if one does not believe that they should be reasoning about $R^2$, they may find themselves actually doing so unknowingly.
Our proposed prior, on the other hand, makes prior knowledge on the $R^2$ of an auto-regressive model explicit, while benefiting from the same shrinkage properties usually sought out in previous work.
The priors suggested in this paper also differ substantially from the default or Jeffreys priors which have been considered for simple time-series model in the past \citep{zellner1996introduction,berger1994noninformative}. \citet{liseo2013objective} recommend not using such priors for AR processes of order larger than 4, so we do not entertain comparison to these priors any further.  

\section{A prior on $R^2$ for auto-regressive models}\label{sec:arr2-prior}
In the following, we will denote the population variance operator $\var{x}$ by $\sigma^2_{x}$, its conditional variant $\var{x\mid\theta}$  by $\sigma^2_{x\mid \theta}$, and their unbiased sample estimators by $\hat\sigma^2_{x}$ and $ \hat{\sigma}^2_{x\mid \theta}$ respectively. Further, we denote by $\theta$ the vector of all parameters entering the observation model. Say, $y_t = \phi y_{t-1} + \epsilon_t$ and $\epsilon_t \sim \normal(0,\sigma^2)$, assuming weak stationarity, then $\theta=(\phi,\sigma^2)$, and $\sigma^2_{y|\theta} = \frac{\sigma^2}{1-\phi^2}$. Further, we suppress the conditioning set of parameters and data when using the ``$\sim$'' symbol for readability.
\subsection{The coefficient of determination in AR models}
Consider a pure auto-regressive time-series model of order $p$, denoted by $\ar(p)$, as follows
\begin{equation}
    y_t = \sum_{i=1}^{p} \phi_i y_{t-i} + \epsilon_t, \label{eq:ar}
\end{equation}
where $y_t$ is the observed value at time $t$, $\phi_i$ are the regression coefficients, and $\epsilon_t$ is white noise with variance $\sigma^2>0$. In the following, denote by $\mu_t$ the term of the linear model independent of the contemporaneous white noise term, $\mu_t=\sum_{i=1}^p \phi_i y_{t-i}$.

Assume that $\phi$ is in the stationary region $\phi \in M^{p \times 1} \subseteq \mathbb{R}^{p\times1}$ such that the roots of the characteristic polynomial $\phi(u) = 1 -\phi_1u-\dotsc-\phi_pu^{p}, u \in \mathbb{C} $ lie outside the unit circle  \citep{hamilton2020time}. This implies a stable, weakly stationary time-series with $\vert\E{y_t}\vert < \infty$ and $\sigma_{y_t}^2 < \infty$. 
Without loss of generality, we will assume throughout that the time-series process is centred on zero. 
The fraction of variance that cannot be explained by our model right before time-point $t$ is $\sigma^2/\var{y_t}$. Similar to  \citet{nelson_interpretation_1976}, we  define $R^2$ for auto-regressive models as
\begin{equation} \label{eq:General_R2_def}
    R^2 = 1 - \frac{\sigma^2}{\var{y_t}} = \frac{\sigma^2_{\mu_t}}{\sigma^2_{\mu_t} + \sigma^2},
\end{equation}
where the final equality stems from the white noise definition, implying that current white noise and lags are independent. Hence, setting a prior on the model parameters coherent with a prior on $R^2$  necessitates a joint prior on the variance of the predictor term and of the noise variance.  
\subsection{Prior derivation} \label{sec:ar_prior_derivation}
We now derive the hierarchies and distributional assumptions needed on the model parameters in order to induce a $\betadist(\mu_{R^2},\varphi_{R^2})$ for $R^2$. Throughout, we parameterise the beta distribution in terms of a location $\mu_{R^2}$ and precision $\varphi_{R^2}$ for convenience of interpretation~\citep[see e.g.,][]{aguilar_intuitive_2023}. Consider  a multivariate normal prior centred on the zero vector with covariance $\Lambda = \text{diag}(\lambda_1^2,\dots,\lambda_p^2)$
\begin{equation} \label{eq:basic_prior}
    \phi_i \sim \normal(0, \lambda_i^2), \quad i = 1,\dots,p.
\end{equation}
Denote by $y_{-p} = (y_{t-1} \cdots y_{t-p})^{\prime}$ the vector of $p$ lagged observations and by $\theta$ a vector of all parameters of the observation model $\theta = (\phi,\sigma^2)$. 
Then, the variance expression of the predictor term is
\begin{IEEEeqnarray}{rl}
\sigma^2_{\mu_t} \;& = \E{\var{y_{-p}^{\prime}\phi\mid y_{-p}}} + \var{\E{y_{-p}^{\prime}\phi\mid y_{-p}^{\prime}}} \label{eq:marginal-variance-explained}\\
        \;& = \tr{\Lambda\Sigma_y} \label{eq:marginal-variance-explained-1} \\
        \;& = \sum_{i=1}^p\lambda_i^2(\E{y_{t-i}y_{t-i}|\theta} + \sigma^2), \label{eq:marginal-variance-explained-2}
\end{IEEEeqnarray}
where $\Sigma_y$ is the lag covariance matrix with diagonal $(\E{y_{t-1}y_{t-1}\vert\theta},\dotsc,\E{y_{t-p}y_{t-p}\vert\theta})$. Equation~\ref{eq:marginal-variance-explained}~follows from the law of total variance and the second term vanishes due to the prior mean of zero. Equation~\ref{eq:marginal-variance-explained-1}~follows from the fact that the variance term is a scalar. 
Due to the assumption of weak stationarity of the AR process, the conditional expectation functions are the same for all lags $i \in \{1,\dotsc,p \}$ and given by
\begin{equation}
 \E{y_{t-i}y_{t-i}\vert\theta} = \sigma^2_{y_t\mid\theta} = \phi_1\gamma(1) + \cdots + \phi_p\gamma(p) + \sigma^2 = \phi^{\prime}\gamma + \sigma^2. \label{eq:data-var}
\end{equation}
The auto-covariance function is $\gamma(k) = \text{cov}(y_t, y_{t-k}\vert \theta) = \E{y_{t}y_{t-k} \mid \theta}$, $\{\gamma(k) \in \mathbb{R} : k \in \mathbb{Z}\}$, where $k$ stands for the order. Again, due to the weak stationarity assumption, these are only functions of $k$ (the time distance between lags) and not $t$ (the time index itself).\footnote{The set of all auto-covariances are described by a set of homogeneous difference equations, whose system is named after \citet{yule_vii_1927} and \citet{walker_periodicity_1931}.} The total variance of the predictor term in Equation~\ref{eq:marginal-variance-explained-2} is therefore the scaled sum of prior variances
\begin{equation}
    \sigma^2_{\mu_t} = \sigma^2_{y_t\vert \theta}\sum_{i=1}^p\lambda_i^2 = \sigma^2_{y_t\vert \theta} \tau^2. \label{eq:marginal-variance-explained-3}
\end{equation}
The $R^2$ in Equation~\ref{eq:General_R2_def} reduces to
\begin{equation} \label{eq:margR2_def}
    R^2 =\frac{\sigma^2_{y_t\mid \theta}\tau^2}{\sigma^2_{y_t\mid \theta}\tau^2 + \sigma^2} = \frac{\tau^2}{\tau^2 + \sigma^2 / \sigma^2_{y_t\mid\theta}}.
\end{equation}
By a change of variables, a beta prior on $R^2$ then implies a generalised beta-prime distribution~\citep{johnson1995continuous} for the sum of scaled prior variances of the $\ar$ coefficients
\begin{equation}
    \tau^2 \sim \text{GBP}\left(\mu_{R^2},\,\varphi_{R^2},\,1,\,\frac{\sigma^2}{\sigma^2_{y_t\mid \theta}}\right).
\end{equation}
The GBP distribution can be obtained through a transformation of a beta-prime distributed variable, so if $x \sim \text{BP}(a,b)$, then $y = d x^{1/c} \sim \text{GBP}(a,b,c,d)$ for $a,b,c,d > 0$ \citep{johnson1995continuous}. Therefore, an equivalent prior can be found by scaling the prior variance of $\phi_i$ by $\sigma^2 / \sigma^2_{y_t\mid \theta}$, implying $R^2 = \tau^2/(\tau^2+1)$ and the prior for $\tau^2$ reduces to the simpler beta-prime distribution~\citep{johnson1995continuous},  $\tau^2 \sim \text{BP}(\mu_{R^2},\varphi_{R^2})$. 
Hence, prior~\ref{eq:basic_prior} needs to be scaled by $\sigma^2/\sigma^2_{y_t}$ in order for the implied prior in $R^2$ to remain non-constant with respect to changes in $\sigma^2$ or $\sigma^2_{y_t \mid \theta}$. 

In a final step, we allow for adaptivity of shrinkage at the coefficient level, as per the global-local prior framework \citep{polson2010shrink}, and decompose the sum of prior scales as
\begin{equation}
    \lambda_i^2 = \tau^2\psi_i,
\end{equation}
with $\psi_i\geq0$ lying on the probabilistic simplex so that $\sum_{i=1}^p\psi_i = 1$. We posit the natural Dirichlet prior over the vector $\psi$ following \citet{zhang_bayesian_2022}. Each $\psi_i$ determines the fraction of total variance allocated to the $i\text{th}$ lag. 
We summarise the model hierarchy in Definition~\ref{prop:arr2}.

\begin{definition}\label{prop:arr2}
    The ARR2 prior defined over an auto-regressive model of order $p$ yields the following model structure
    \begin{IEEEeqnarray}{rl}
        y_t \;& \sim\normal (\mu_t,\sigma^2),\quad t=p+1,\dotsc,T\label{eq:arr2_likelihood} \\
        \mu_t \;&= \sum_{i=1}^p\phi_iy_{t-i}\label{eq:arr2_linear_predictor}  \\
        \phi_i \;&\sim \normal\left(0, \frac{\sigma^2}{\sigma_{y_{t}\mid \theta}^2}\tau^2\psi_i\right)\label{eq:arr2_phi_prior} \\
        \tau^2 \;&= \frac{R^2}{1 - R^2}\label{eq:arr2_r2_transform} \\ 
        R^2\;&\sim \betadist(\mu_{R^2},\varphi_{R^2})\label{eq:arr2_r2_prior} \\ 
        \sigma^2\;&\sim \pi(\sigma^2)\label{eq:arr2_sigma_prior} \\
        \psi\;&\sim \Dirichlet(\xi_1,\dotsc,\xi_p).\label{eq:arr2_dirichlet}
    \end{IEEEeqnarray}
\end{definition}
To make inference on the total variance term $\tau^2$, we formulate the equivalent prior in $R^2$ space and transform back appropriately via Equation~\ref{eq:arr2_r2_transform}.

To avoid $\sigma^2_{y_t \mid \theta}$ being dependent on parameters since this can complicate the posterior geometry, we propose the simplification of replacing $\sigma^2_{y_t \mid \theta}$ with the single data-based estimate $\hat{\sigma}^2_{y}$. This is common practice within the previous literature for $R^2$ priors \citep{aguilar_intuitive_2023,aguilar2024generalized}. In fact, $\hat{\sigma}^2_{y}$ can be shown to be a conservative estimate of the conditional variance, since $ \sigma^2_{y} = \mathbb{E}[\sigma^2_{y_t\vert \theta}]+ \var{\mathbb{E}[y_t\vert \theta]}$ $\implies \sigma^2_{y}\geq \mathbb{E}[\sigma^2_{y_t\vert \theta}]$. From Equation~\ref{eq:margR2_def} one can see that this will tend to bias $R^2$ slightly upwards. Yet, shrinkage properties of the prior on $\phi$, that is, the behaviour of the prior near the origin and the tails \citep{polson2010shrink}, remain unaltered by this modification. These are controlled by the GBP parameters $ac$ and $bc$ respectively \citep{yanchenko_r2d2_2021}. Additionally, we expect the data-based estimate to be reasonable, as the term $\E{y_t\vert \theta}$ as defined by Equation~\ref{eq:ar} is 0.
\subsection{Specifying hyperparameters and connections to other priors} \label{sec:comp_alt_priors}

For the ARR2 prior, the hyperparameters controlling the beta prior on the $R^2$ and the Dirichlet prior on $\psi$ need to be specified by the modeller. The prior on the $R^2$ encodes beliefs about the variance explained by the model, while the prior on $\psi$ encodes beliefs about the contribution of each lag to the explained variance, and the degree of sparsity. Their hyperparameters alter the  prior's properties for $\phi$ at the origin as well as the tails \citep{zhang_bayesian_2022}. \citet{zhang_bayesian_2022} use a uniform prior on $R^2$, induced by $(\mu_{R^2}, \varphi_{R^2})$ = (0.5, 2), while \citet{aguilar_intuitive_2023} instead recommend $ (\mu_{R^2},\varphi_{R^2})=(0.5,1)$. They show that setting $(\mu_{R^2},\varphi_{R^2})$ such that $(1-\mu_{R^2})\varphi_{R^2} \in [0,0.5]$ induces fat enough tails so that the marginal prior for $\phi$ is of bounded influence, meaning not shrinking sufficiently large signals. This additionally causes the $R^2$ prior to exhibit a bathtub-like shape with relatively high mass at 0 and 1. Large mass at 1, however, is undesirable for stationary AR models, since the stationary region for $\phi$ implies $R^2 < 1$\footnote{Suppose $\phi \in M^{p\times 1}$ and $\sigma^2 > 0$. Then, by the properties of weak stationarity, $\text{var}(y_{-p}^{\prime}\phi) < \infty$. This renders $\text{var}(y_{-p}^{\prime}\phi)/ (\text{var}(y_{-p}^{\prime}\phi) + \sigma^2) < 1$.}. It is important to recognise that non-stationary processes may occur even when \(R^2 < 1\), particularly in finite samples, and the decomposition defined by the Dirichlet prior will also impact the prior probability of stationarity. Supplementary Material Section A shows that for the ARR2 prior, the majority of the prior mass lies in the stationary region, however this can be influenced by the hyperparameters of the \(R^2\) prior and its simplex decomposition. Decompositions that heavily regularise higher order lag polynomials to zero reduce the probability of prior non-stationarity since lower dimensional $\ar$~models have less complex restrictions on the stationary parameter space \citep{huerta1999priors,heaps_enforcing_2022}.

As a default, we prefer $R^2 \sim \betadist(1/3,3)$ which exerts shrinkage toward lower $R^2$ values, yet has a small, nearly constant gradient. This implies a long tail with reasonable mass on larger $R^2$ values. 
Based on our experiments, this is a sensible default, but potential prior-likelihood conflict may be checked, for example, with power-scaling sensitivity analysis~\citep{kallioinenDetectingDiagnosingPrior2023}.Supplementary Material Section B shows a plot of our recommended $R^2$ prior along with two alternatives.

As discussed by \citet{zhang_bayesian_2022}, the larger the prior concentrations $\xi_i$, the more the Dirichlet distribution will resemble a uniform $\psi \approx (1/p,\dots,1/p)$, whereas small concentration values lead to more mass in the edges of the simplex, inducing stronger sparsity patterns in $\phi$. \citet{aguilar_intuitive_2023} recommends setting $\xi_i=0.5$ to encourage sparse posteriors\footnote{In practice, we have found $\xi_i =1$ to  yield clear sparsity patterns while remaining computationally feasible.}.  By setting these hyperparameters appropriately, the variance decompositions can resemble priors commonly used in time-series modelling (Table~\ref{tab:ar-priors}). Importantly, unlike these other priors, the ARR2 allows independently changing the prior on the $R^2$ without affecting this decomposition (see Figure~\ref{fig:r2-contribution}). We consider two types of decompositions motivated by priors popular in time-series analysis, discussed below.

\begin{figure}[!t]
    \centering
    \input{tikz/r2_plot_priors}
    \caption{Implied prior $R^2$ for an AR model with 12 lags. The top plot shows the prior $R^2$ induced by the \textcolor{gray}{ARR2 prior (grey)} with our suggested hyperparameters for the \(R^2\) prior, \textcolor{minnesotacolour}{Minnesota-type prior (yellow)}, \textcolor{gaussiancolour}{independent Gaussian priors (green)}, and the \textcolor{rhscolour}{regularised horseshoe prior (purple)} For simplicity, we assume marginal variance of $y_t$ and $\sigma$ equal to 1. The bottom plots show the prior means of the relative contributions to the $R^2$ of the regression coefficients. Here the different hyperparameter settings for the ARR2 are shown: \textcolor{arr2flatcolour}{flat (blue)}, \textcolor{arr2minncolour}{Minnesota-type (red)}, \textcolor{arr2bumpcolour}{bump (light blue)}. The ARR2 (flat), RHS and Gaussian lines are overlapping as they all have a flat decomposition structure.}
    \label{fig:r2-contribution}
\end{figure}

The Minnesota prior, often used for economic data, was originally proposed by \citet{doan1984forecasting} and is motivated by the finding that stationary time-series exhibit auto-correlations which are often well described by exponential decay. We consider the hierarchical version presented by \citet{carriero_bayesian_2015} which may be viewed as a non-hierarchical normal-gamma prior \citep{brown2010inference}. 
Increasing shrinkage with each lag polynomial favours initial lag components to contribute most to total variance.

The regularised horseshoe prior~\citep[RHS;][]{piironen_hyperprior_2017} on the other hand is a more general sparsity-inducing prior which has been recently adopted for many time-series models in economics~\citep{chan_minnesota-type_2021,huber2021inducing,kohns2024horseshoe}.
It belongs to the class of global-local priors~\citep{polson_half-cauchy_2012} in which fat tails favour either shrinking a coefficient strongly toward zero, or only very little.
Following \citet{piironen_hyperprior_2017}, we scale the global scale $\tau_{\text{RHS}}$ by the hyperparameter $\tau_0$, defined by a prior on number of active coefficients (set here to half the number of lags included in the model). Additionally, a default choice of independent Gaussians with unit variance is commonly used as a weakly informative prior in linear regression models, and we compare to this as a baseline.

\begin{table}[t]
    \centering
        \small
    \begin{tabular}{lll}
    \toprule
         & $\pi(\phi_i)$ & Scales\\
        \midrule
        Minnesota & $\normal\!\left(0,\kappa/i^2\right)$ & $\kappa \sim \gammadist(1,1/0.04)$ \\
        Regularised horseshoe & $\normal\!\left(0, \tau_{\text{RHS}}^2\tilde{\lambda}^2_i\right)$ 
 & $\begin{cases}
            \tilde{\lambda}_j^2= \frac{c^2\lambda_j^2}{c^2 + \tau^2\lambda_j^2}, \lambda_j\sim \Cauchy_+(0,1) \\ 
            \tau_{\text{RHS}} \sim \Cauchy_+(0,\tau_0)
        \end{cases}$ \\
                Gaussian & $\normal\!\left(0,1\right)$ & -- \\
        \bottomrule
    \end{tabular}
    \caption{Specifications of alternative priors. 
    }
    \label{tab:ar-priors}
\end{table}

Figure~\ref{fig:r2-contribution} shows in the upper panel the induced prior for $R^2$, and in the lower, relative $R^2$ contributions\footnote{Conditional on the parameters, the conditional $R^2$ for the $i\text{th}$~lag can also be shown to reduce to the squared $i\text{th}$-degree partial auto-correlation (see Supplementary Material Section F.2). In the frequentist treatment this is used to choose the appropriate AR order~\citep{box_time_1994}.} from each lag polynomial which we denote by
\begin{equation}
    R^2_i:= \frac{\var{\phi_iy_{t-i}}}{\var{y_t}}.
\end{equation}
As shown, with appropriate hyperparameter settings, the ARR2 prior can resemble the other priors, or can encode different beliefs entirely. In the simulated data experiments (Section~\ref{sec:simulations}) and case study (Section~\ref{sec:case-study}), we will set the prior concentration values to generate comparable shapes to the Minnesota ($\xi_i = (p^2/10 \cdot 1/i^2)$) and RHS ($\xi = (0.1,\dots,0.1)$).\footnote{Scaling the concentration values for the Minnesota ARR2 by $p^2/10$ enforces a lower bound of 0.1, which we have found empirically to be a threshold under which computational issues arise with sampling algorithms.}

The previous discussion has focused on the prior properties in $R^2$-space. It is also possible to analyse the induced priors for the AR coefficients, the partial autocorrelations of the AR process, and the roots of the characteristic polynomial. This is shown in Supplementary Material Section A. Also in these spaces, the ARR2 (Minn.) mostly resembles the Minnesota prior and ARR2 (flat) the RHS, respectively.

For simplicity, such analyses consider the observation noise variance to be known and fixed. With respect to the AR coefficients, we conjecture that the marginal properties of our prior are similar to those discussed in \citet{aguilar_intuitive_2023} under the additional assumption of fixing $\sigma^2_{y_t \mid \theta}$ to a known scalar. We refer the reader to  \citet{aguilar_intuitive_2023}'s work for further comparison to other popular shrinkage priors.

\FloatBarrier

\subsection{Auto-regressive models with exogenous covariates}\label{sec:arr2-for-arx}
Auto-regressive models with exogenous covariates (ARX models) are an extension to pure AR models which allow the modeller to incorporate exogenous information (independent of the lagged covariates). They are used widely in economic analysis as building blocks for multivariate extensions to AR models~\citep{litterman1986forecasting}, and for monitoring in various fields including building and structure engineering~\citep{saitoBayesianModelSelection2010,matsuokaBayesianEstimationInstantaneous2021,matsuokaBayesianTimeFrequency2020,kimLongtermBridgeHealth2018,barraza-barrazaAdaptiveARXModel2017}, medicine~\citep{fangBayesianInferenceFederated2021,nunesARXModelingDrug2013}, and environmental sciences~\citep{zanottiChoosingLinearNonlinear2019,zhangEstimatingDynamicSolar2022}. ARX models are defined formally as
\begin{equation}
    y_t = \sum_{i=1}^{p} \phi_i y_{t-i} + x_t^{\prime}\beta + \epsilon_t, \label{eq:arx}
\end{equation}
where $X \in \mathbb{R}^{T\times m}$ is the exogenous design matrix, $\beta$ is an $m\times 1$ vector of parameters and $\epsilon_t$ is some white-noise term with variance $\sigma^2$.

Here, the predictor term $\mu_t = \sum_{i=1}^p\phi_iy_{t-i} + x_t^{\prime}\beta$, now defines the total variance $\sigma^2_{\mu_t} = \var{\phi^{\prime}y_{-p}} + \var{x_t^{\prime} \beta}$ which we can now decompose as in Section~\ref{sec:ar_prior_derivation}. 
Assume that the covariates $X$ are scaled to 0 mean and unit variance. 
Since $\E{x_t^{\prime}\beta y_{t-k}} = 0$ for all lag polynomials, it is easy to verify that $\var{\phi^{\prime} y_{-p}}$ induces the same set of conditional variance functions as a pure AR model (Supplementary Material Section C). 
From this, we can repeat the probabilistic arguments of Section~\ref{sec:arr2-prior} to form an $R^2$ prior on ARX models (which we will refer to as the ARR2 prior for ARX models).
\begin{definition}\label{prop:arxr2}
    The ARR2 prior defined over an $\arx(p,m)$ model yields the following model structure
    \begin{IEEEeqnarray}{rl}
        y_t \;& \sim\normal (\mu_t,\sigma^2),\quad t=p+1,\dotsc,T \\
        \mu_t\;&= \sum_{i=1}^p\phi_iy_{t-i} + x_t^{\prime}\beta \\
        \phi_i \;&\sim \normal\left(0, \frac{\sigma^2}{\sigma_{y_{t}\mid \theta}^2}\tau^2\psi_i\right) \\
        \beta_j \;&\sim \normal\left(0, \frac{\sigma^2}{\sigma_{x_j}^2}\tau^2\psi_{j}\right) \\
        \tau^2 \;&= \frac{R^2}{1 - R^2} \\
        R^2\;&\sim \betadist(\mu_{R^2},\varphi_{R^2}) \\ 
        \sigma^2\;&\sim \pi(\sigma^2) \\
        \psi\;&\sim \Dirichlet(\underbrace{\xi_1,\dotsc,\xi_p}_{i=1,\dotsc,p}, \underbrace{\xi_{p+1}, \dotsc,\xi_{p+m}}_{j=1,\dotsc,m}),
    \end{IEEEeqnarray}
\end{definition}
A more complete derivation of this prior is provided in Supplementary Material Section E.%~\ref{appendix:arxr2-derivation}.
This prior makes explicit that both the $\ar$ and exogenous regression components contribute to the total variance of the model $\tau^2$, which is then decomposed probabilistically via $\psi$. In addition to the $p$ components for each lag, $\psi$ now includes $m$ additional components attributed to each exogenous covariate. 
These two sets of components compete for the relative contribution to the total variance, and the hyperparameters $\xi$ can be used to nudge the model toward particular decompositions.

The hyperparameters $\left(\xi_1,\dotsc,\xi_p\right)$ may be set according to the proposed time-series lag structure, as in Figure~\ref{fig:r2-contribution}, and $\left(\xi_{p+1},\dotsc,\xi_m\right)$ according to prior knowledge of sparsity and correlation structure. For example, one may set $(\xi_1,\dotsc,\xi_p)$ according to Minnesota decay as for the AR models, while imposing sparsity-favouring decomponsition for $x_t$ by setting $(\xi_{p+1},\dotsc, \xi_{p+m})$ to a relatively low value of 0.1.
\subsection{State-space models}
\label{sec:arr2_derivation_state_space_models}
Similar to the derived $R^2$ prior for observable time-series dynamics, we present in this section an extension to the state-space framework which is a flexible family of models for joint estimation of observable and latent time-series. These have a long tradition in Bayesian estimation~\citep{kitagawa1984smoothness,west2006bayesian} and probabilistic filtering~\citep{sarkka2023bayesian} and are routinely used in the social and hard sciences~\citep{chan2023bayesian}. 

We consider state-space models of the following form: 
\begin{IEEEeqnarray}{rlrl} 
    y_t  \;& = x_t^{\prime}\beta + s_t^{\prime}G + \epsilon_t,&\quad \epsilon_t \;&\sim \normal(0,\sigma^2) \label{eq:gen_state_space_observation} \\
    s_t \;& = \Phi s_{t-1} + e_t,& e_t \;&\sim \normal(0,\Sigma_s) \label{eq:gen_state_space_transition},
\end{IEEEeqnarray}
where $y_t \in \mathbb{R}$ is the scalar valued target and $x_t \in \mathbb{R}^{m\times 1}$ exogenous covariates. 
$s_t \in \mathbb{R}^{Q\times 1}$ are unknown states, $G$ is a state coefficient vector, $\Phi \in \mathbb{R}^{Q\times Q}$ is some invertible state transition matrix, and $\Sigma_s$ the state error covariance matrix with diagonal elements $(\sigma_{s_1} ,\dots,\sigma_{s_Q})$. 
Assume, further, that there are $Q$ initial conditions to the states, $s_0 \sim N(0,\Sigma_s)$. 
The equation for $y_t$ is called the observation equation in the state-space literature and that for $s_t$ the transition equation~\citep{harvey_forecasting_1990}. 
The generality of the state-space model in Equations~\ref{eq:gen_state_space_observation}-\ref{eq:gen_state_space_transition} can be seen from the fact that AR, MA, ARMA, ARX, Bayesian structural time-series models~\citep{brodersen2015inferring,scott2014predicting}, and many more, are special cases~\citep{harvey_forecasting_1990}.\footnote{While Equation~\ref{eq:gen_state_space_transition} is formulated as an $\ar(1)$ process, any scalar or vector valued auto-regressive time-series of order $p$ can be written as an $\ar(1)$ process via its companion form~\citep{hamilton2020time}.} 

Define the predictor term of the observation equation as $\mu_t = x_t^{\prime}\beta + s_t^{\prime}G$.
Under the assumption of independence between $x_t$ and states, the variance of the observation equation can be factored as $\sigma^2_{\mu_t} = \var{x_t^{\prime} \beta} + \var{s_t^{\prime}G}$. 
As before, the variance contribution of the exogenous covariate component simplifies to $\var{x_t^{\prime} \beta} = \sum_{j=1}^m\lambda_j^2\sigma_{x_j}^2$. 
The marginal variance contribution of the states is complicated by the law of motion in Equation~\ref{eq:gen_state_space_transition}, however, we make three simplifying assumptions for ease of exposition of the following defintion:
\begin{enumerate}[nosep]
    \item $\Phi = \mathrm{diag}(\phi_1,\dots,\phi_Q)$ (each state follows an $\ar(1)$ process); \label{ass1}
    \item $G = \mathbf{1}_Q$ (coefficients on state are a $Q$-dimensional vector of ones); \label{ass2}
    \item and, roots of $ 0 = \mathrm{det}(\boldsymbol{I_{Q}} - \Phi u), u\in\mathbb{C}$, are within the unit circle (states are stationary), \label{ass3}
\end{enumerate}
where $\boldsymbol{I_{Q}}$ denotes the $Q$-dimensional identity matrix.
While assumptions~\ref{ass1} and~\ref{ass2} may seem limiting for the prior on a state-space model's $R^2$, a large class of models such as dynamic regressions~\citep{chan2020reducing}, unobserved component models in economics~\citep{fruhwirth2010stochastic} and forecasting models~\citep{kohns_nowcasting_2022,potjagailo2023flexible} similarly assume independent $\ar(1)$ processes for the states. 
In Supplementary Material Section F,%~\ref{appendix:state_space_derivations},
we derive the $R^2$ relaxing some of these assumptions. 
Assumption~\ref{ass3} is needed in order for the $R^2$ to be well defined, as $R^2 \rightarrow 1$ with non-stationarity of the states.

\begin{definition}\label{prop:ssr3}
    The ARR2 prior for discrete state-space models of form in Equations~\ref{eq:gen_state_space_observation}--\ref{eq:gen_state_space_transition} under Assumptions~\ref{ass1}--\ref{ass3} yields the following model structure
    \begin{IEEEeqnarray}{rl}
        y_t \;& \sim\normal (\mu_t,\sigma^2),\quad t=1,\dotsc,T \\
        \mu_t \;&= x_t^{\prime}\beta + s_t^{\prime}\mathbf{1}_{Q} \\
        s_{t,q} \;&\sim \normal\left(\phi_q s_{t-1,q},\, \sigma^2 \tau^2\left(1-\phi_q^2\right) \psi_q \right) \label{eq:prior_states} \\ 
        s_{0,q} \;&\sim \normal\left(0,\,\sigma^2\tau^2 \left(1-\phi_q^2\right)\psi_q \right) \\
        \phi_q \;&\sim \normal\left(0,\sigma_{\phi_q}^2\right), \quad q = 1,\dotsc,Q \\
        \beta_j \;&\sim \normal\left(0, \frac{\sigma^2}{\sigma_{x_j}^2}\tau^2\psi_{j}\right) \\
        \tau^2 \;&= \frac{R^2}{1 - R^2} \\
        R^2\;&\sim \betadist(\mu_{R^2},\varphi_{R^2}) \\ 
        \sigma^2\;&\sim \pi(\sigma^2) \\
        \psi\;&\sim \Dirichlet(\underbrace{\xi_1,\dotsc,\xi_Q}_{q=1,\dotsc,Q}, \underbrace{\xi_{Q+1}, \dotsc,\xi_{m}}_{j=1,\dotsc,m}).
    \end{IEEEeqnarray}
\end{definition}
A more complete derivation of this prior is provided in Supplementary Material Section F.
%~\ref{appendix:state_space_derivations}.
%
The hierarchy of the ARR2 prior makes three types of dependencies explicit. 
First, unlike previous priors for state-space models, the states for all $t$ depend on the noise variance, $\sigma^2$. 
This is needed in order for the relative shares of explained variance between the regression and state component to stay constant to changes in the scale of the observation noise. 
One may think of this as a state component analog to the prior scaling of $\beta$ by $\sigma^2$ in Definition~\ref{prop:arxr2}. 
Second, a state's prior variance decreases with its AR parameter $\phi_q$. This makes intuitive sense, since not much can be learned from states which are random noise with low serial correlation. Lastly, the ARR2 prior makes joint shrinkage of the state and regression component explicit via decomposition of the total variance $\tau^2$. 
Previous approaches on the other hand, assume independence between the state and other model components in the observation equation~\citep{cadonna2020triple}. This neglects the fact that  states and other components in the observation equation compete to explain shares in the total variance of $y_t$.

The dimensionality of the estimation problem in Definition~\ref{prop:ssr3} can easily get very high. 
Consider the addition of $p$ lags per covariate.
In these situations, one can reduce computational complexity by decomposing the prior variance at the group-level: $\beta_{i,j} \sim \normal(0,\tilde{\psi}_{i,j}\tau^2) $, where $\tilde{\psi}_{i,j} = w_j \psi_i $, $\psi \sim \Dirichlet(\xi_1,\dots,\xi_m)$ and $\sum_{j=1}^p w_j = 1$ are some deterministic weights for $j \in \{1,\dots,p\}$ and $i \in \{1,\dots,m\}$. In its simplest form, set $w_j=1/p \forall j$ such that the simplex dimensionality reduces to $m+Q$ instead of $mp + Q$. 
Inspired by a Minnesota-like decomposition for a group of lags, one may then set $w_j$ to $(1/j^2) / (\sum_{s =1 }^p 1/s^2)$. 
\paragraph{Local-linear trend model with covariates}
To illustrate the above, consider the following simplified Bayesian structural equation model, referred to as the local-trend model (LTX):
\label{eq:llt_def}
\begin{IEEEeqnarray}{rlrl} 
    y_t  \;& = x_t^{\prime}\beta + \delta_t + \epsilon_t,&\quad \epsilon_t \;&\sim \normal(0,\sigma^2) \\
    \delta_t \;& = \phi \delta_{t-1} + e_t,& e_t \;&\sim \normal(0,\sigma_{\delta}^2),
\end{IEEEeqnarray}
where $\delta_t$ is an unobserved time-trend and $\phi \in (-1,1)$. 
It can be shown that Equation~$\ref{eq:llt_def}$ implies an ARMA-X model for $y_t$~\citep{harvey_forecasting_1990}. 
The $R^2$ for this prior is then defined as:
\begin{equation} \label{eq:llt_r2_def}
   R^2 = \frac{\overbrace{ \sigma^2\sum_{j=1}^m\lambda_j^2 + \sigma^2 \frac{\sigma_{\delta}^2}{1-\phi^2} }^{\coloneqq \sigma^2 \tau^2}}{\sigma^2 \tau^2 + \sigma^2} = \frac{\tau^2}{\tau^2 + 1}.
\end{equation}
Hence, $R^2 \sim \betadist(\mu_{R^2},\varphi_{R^2})$ again implies $\tau^2 \sim \betaprime(\mu_{R^2},\varphi_{R^2})$. 
Then 
\begin{IEEEeqnarray}{rlrl} 
        \delta_t & \sim \normal(\phi \delta_{t-1},\sigma^2(1-\phi^2)\psi_{1}\tau^2), \\
        \delta_{0} & \sim \normal(0,\sigma^2(1-\phi^2)\psi_{1}\tau^2) \\
        \beta_j & \sim \normal(0,\sigma^2/\sigma^2_{x_j}\psi_{j+1}\tau^2),\quad j = 1,\dots,m.
\end{IEEEeqnarray}
%. 
%
In Supplementary Material Section F.1,
we show how the ARR2 prior can be applied to the popular dynamic regression model.
\section{Simulated data experiments}\label{sec:simulations}
In this section, we primarily aim to understand the quality of parameter estimation and predictions when estimating models of increasing complexity for different data-generating processes (DGPs) simulated from AR, ARX and LTX models. 
We investigate the behaviour when the complexity of the DGP stays fixed as the complexity of the estimated model grows. 
This is a common problem in the Bayesian workflow~\citep{gelman2020bayesian} in which model building will often involve sequentially adding more complexity. 
For all simulations, we fit the proposed ARR2 prior with two hyperparameter settings, Minnesota ARR2 and flat ARR2 (see Section~\ref{sec:comp_alt_priors}). We compare these models to two types of commonly used prior for time-series analysis, the Minnesota~\citep{giannone_prior_2012} and regularised horseshoe~\citep[RHS;][]{piironen_hyperprior_2017}, as well as a relatively wide non-hierarchical 
standard Gaussian prior. For all simulation experiments, we simulate 25 sets of data from each DGP. 

We measure goodness of parameter recovery by average root-mean-squared error ($\mathrm{RMSE}$) between the posterior mean and the true coefficients. Let $\hat{\theta} = 1/S
\sum^S_{s = 1 }\theta^{(s)}$, where $S$ are the number of retained posterior draws. We define $\mathrm{RMSE}$ as: 
\begin{equation}
    \mathrm{RMSE} \coloneqq \sqrt{1/K|| \hat{\theta} - \theta||_2^2},
\end{equation}
where $K$ is the dimensionality of $\theta$ and $||\theta||^2_2$ calculates the squared Euclidean norm. %and $\hat{\theta}$ is the posterior mean based on the a simulated data set.

Predictions are evaluated using the leave-future-out (LFO) expected log predictive density~\citep[elpd;][]{vehtari_practical_2017,burkner_approximate_2020}, for $M$--step-ahead future observations, which we compute as
\begin{equation}
    \mathrm{elpd}_\mathrm{LFO} \coloneqq \sum_{i=L}^{T-M} \log p(y_{i+1:M}\mid y_{1:i}),
\end{equation}
which is estimated by
\begin{equation}
    \widehat{\mathrm{elpd}_{\mathrm{LFO}}} = \sum_{i = L}^{T-M} \log\Big( 1/S \sum_{s = 1}^S p(y_{i+1:M}\mid y_{1:i},\theta^{(s)}) \Big).
\end{equation}
In all simulations below, we focus for simplicity on 1-step-ahead LFO predictions $(M=1)$.
For a set of integers $A\subset\mathbb{Z}$, we denote $y_{-A} = \{y_j: j = 1,\dotsc,n,\,j\notin A\}$. 

In order to compare predictions between simulation exercises with differing time-series lengths, we report mean log predictive density (MLPD) which is \(\mathrm{elpd}_\mathrm{LFO} / (T-L)\).
Predictions are always evaluated on a hold-out set where $L$ is equal to half the number of observations ($T/2$).

All experiments were performed using \proglang{Stan}~\citep{stan} using the \texttt{cmdstanr} interface in \proglang{R}, and the source code is freely available at: \url{https://github.com/n-kall/arr2}. Supplementary Material Section C %~\ref{appendix:stan}
has \proglang{Stan}~\citep{stan} code for the ARR2 prior for AR models.

\subsection{AR models}
We generate data from three different AR processes whose AR coefficients mimic 3 types of processes typically found in stationary time-series models: Minnesota, delayed relevance, and  dampened oscillations. Minnesota type and hump shape (delayed relevance) AR parameter profiles are commonly found in economic time-series \citep{doan1984forecasting} which we adapt from the article by
\citet{mogliani2021bayesian}. These DGPs favour priors such as the Minnesota and RHS prior. Dampened oscillations, on the other hand, are more common to physical applications \citep{west2006bayesian} and are included as an example of complicated auto-covariances with many large coefficients. We expect this DGP to disadvantage priors which assume exchangeability in the AR parameters.  
The parameters are defined in Table~\ref{tab:ar-dgps} and are chosen to induce true $R^2$ of around 0.7.
For all DGPs, we set $\sigma^2 = 1$.  The true lag order is always fixed to eight. Under each DGP, we simulate $T = 120$ observations, and fit AR models of increasing order (up to $p = T / 2 = 60$). 

\begin{figure}[t]
    \centering\input{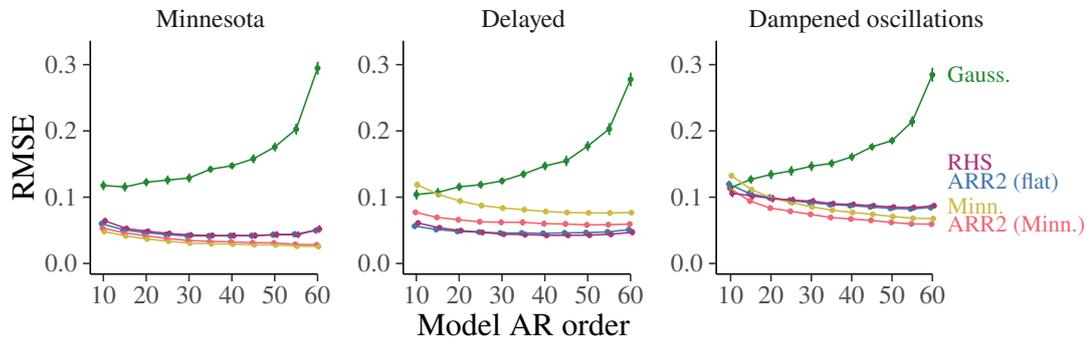}
    \vspace{-\baselineskip}
    \caption{RMSE for $\phi$ in AR simulations with \textcolor{gaussiancolour}{independent Gaussian priors}, a \textcolor{minnesotacolour}{Minnesota-type prior}, the \textcolor{rhscolour}{regularised horseshoe prior}, and our proposed \textcolor{arr2flatcolour}{ARR2 prior with flat concentration} and \textcolor{arr2minncolour}{ARR2 prior with Minnesota concentration}. Means and $\pm$ standard error (of 25 simulations) are shown.
    }
    \label{fig:ar-120-estim}
\end{figure}

\begin{table}[t]
    \centering
        \small
    \begin{tabular}{ll}
    \toprule
        DGP & AR Coefficients\\
        \midrule
        Minnesota& $\phi = (0.6, 0.15, 0.067, 0.038, 0.024, 0.017, 0.012, 0.009)$\\
         Dampened oscillations& $\phi = (-0.509,0.582,-0.069,-0.309,0.242,0.031,-0.166,0.089)$\\
        Delayed relevance& $\phi = (0, 0, 0, 0, 0.7, 0.2, 0.05, 0.025)$\\
        \bottomrule
    \end{tabular}
    \caption{Data generating processes for AR experiments}
    \label{tab:ar-dgps}
\end{table}
The results for the parameter recovery are shown in Figure~\ref{fig:ar-120-estim}. 
As expected, priors whose shrinkage across lag polynomials matches the DGP's coefficient profile perform best. 
The Minnesota ARR2 and Minnesota prior perform best for the DGPs with decaying AR coefficients, while the flat ARR2 and RHS perform best for the delayed DGP.

Interestingly, all priors bar the Gaussian exhibit estimation error which is nearly independent of the lag order. 
Independent Gaussians, result in much higher RMSE than the any of the other priors.
This is particularly noticeable as the model size increases (more irrelevant lags are included). 

\begin{figure}[h!]
    \centering\input{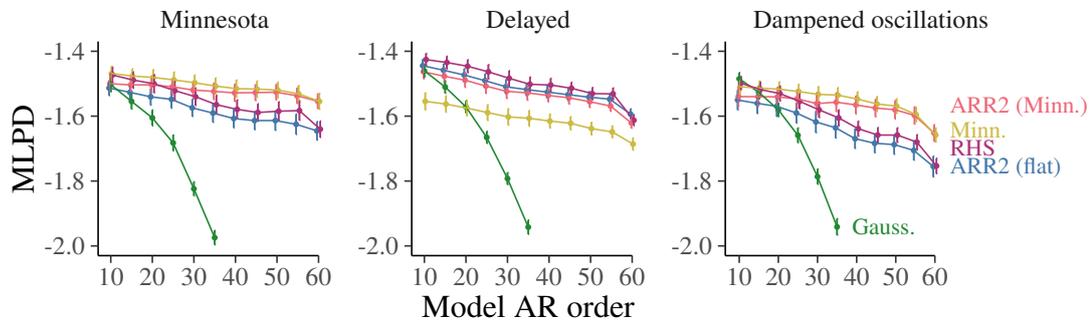}
    \vspace{-\baselineskip}
    \caption{One step ahead leave-future-out mean log predictive density for AR simulations. Posteriors induced by \textcolor{gaussiancolour}{independent Gaussian priors}, a \textcolor{minnesotacolour}{Minnesota-type prior}, the \textcolor{rhscolour}{regularised horseshoe prior}, and our proposed \textcolor{arr2flatcolour}{ARR2 prior with flat concentration} and \textcolor{arr2minncolour}{ARR2 prior with Minnesota concentration}. We sample $T = 120$ observations from each of the respective DGPs and fit AR models of increasing size to them under each of the different priors. Means $\pm$ standard error (of 25 simulations) shown. The MLPD axis is truncated to allow differences between shrinkage priors to be seen.}%
    \label{fig:ar-elpd}
\end{figure}

Figure~\ref{fig:ar-elpd} shows results for the LFO predictive performance. 
These mirror parameter inference performance.
Again, the ARR2, Minnesota and RHS priors are relatively unaffected by model size, indicating that the shrinkage of irrelevant lags works as intended. 
The slight decrease in predictive performance for the largest models is likely due to the fewer in-sample observations available for the larger model sizes (e.g. for the first LFO fold, a model with 60 lags has only one in-sample observation per lag, whereas a model with 9 lags has 6).

Taken together, the ARR2 prior can either match or out-perform alternative priors in both estimative and predictive tasks, all by changing the prior decomposition.

\FloatBarrier

\subsection{ARX models} \label{sec:sim_arx}

For the ARX models, we generate data according to a process
\begin{IEEEeqnarray}{rl}
    x_t \;&\sim \normal(0,\Sigma_X) \\ \nonumber
    y_t  \;&\sim \normal(\sum_{i=1}^p \phi_i y_{t-i} + x_t^{\prime}\beta, \sigma^2), \\ \nonumber
\end{IEEEeqnarray}
where $X$ is an $T \times m$ matrix, and we consider exogenous covariates with $m = \{20, 100, 200, 400\}$.
We follow \citet{piironenComparisonBayesianPredictive2017} by setting the covariance matrix $\Sigma_X$ to be block diagonal, each block of dimension $5 \times 5$. 
Each covariate is standardised and is correlated with the other four covariates in its block with coefficient $\rho=\{0, 0.5, 0.9\}$, and uncorrelated with the covariates in all other blocks. 
Further, the coefficients $\beta$ are such that only the first $15$ covariates influence the target $y$ with coefficients $(\beta^{1:5},\beta^{6:10},\beta^{11:15}) = (\varrho,0.5 \varrho,0.25 \varrho)$ and zero otherwise. We set $\varrho = 0.59$ and $\sigma^2=1$ to set the DGP's $R^2$ to vary between 0.8 and 0.95\footnote{We also follow \citet{piironenComparisonBayesianPredictive2017} in that we adjust $\xi$ so that with increasing correlation, the signal of the $X$ component in the predictor term stays approximately equal.}. 
We set the AR coefficient concentrations according to the Minnesota DGP used for the AR simulation study and those for the exogenous covariates to 0.1. 
Similar DGPs have been also used by \citet{mclatchieRobustEfficientProjection2023}, and \citet{mclatchie_efficient_2023}.

\begin{figure}[!t]
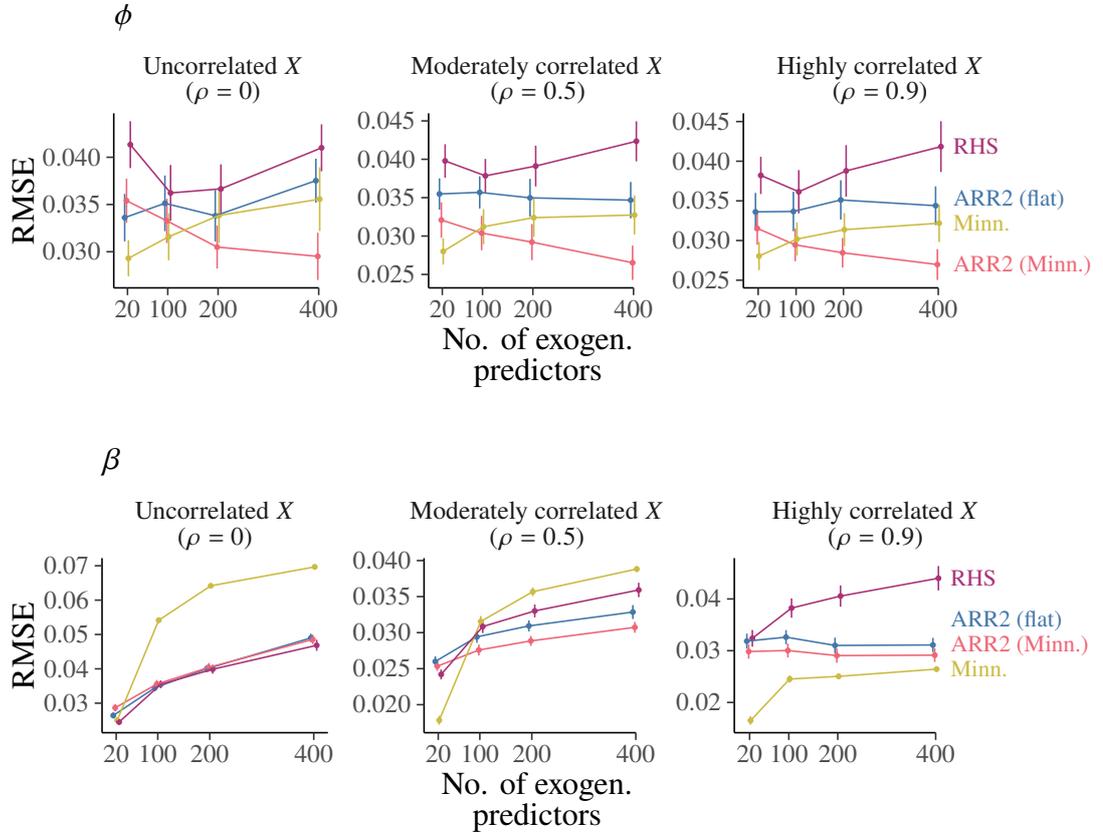


    \begin{minipage}{\textwidth}\centering\input{tikz/arx_estim_rmse_phi}
    \vspace{-2\baselineskip}
    \end{minipage}   
    
    \begin{minipage}{\textwidth}  \centering\input{tikz/arx_estim_rmse_beta}
    \end{minipage}   
    \vspace{-\baselineskip}
    \caption{RMSE of posteriors for ARX simulations induced by  our proposed \textcolor{arr2flatcolour}{ARR2 prior with flat concentration} and \textcolor{arr2minncolour}{ARR2 prior with Minnesota concentration}, a \textcolor{minnesotacolour}{Minnesota-type prior}, and the \textcolor{rhscolour}{regularised horseshoe prior}. Means $\pm$ standard error (from 25 simulations) are shown.}
    \label{fig:arx-estim-rmse}
\end{figure}

We simulate $T = 120$ data points and fit ARX models with 12 AR components (\(p = 12\)) and \(m\) exogenous covariates. 
We specify our proposed ARR2 prior for ARX models with either a flat or Minnesota-type profile for the Dirichlet concentration parameters and compare to Minnesota and RHS.\footnote{Due to poor performance of independent Gaussians exhibited in the AR simulation experiments, it was omitted in the more complex models.} 
The results are shown in Figure~\ref{fig:arx-estim-rmse}.
Here, we examine the parameter inference performance for the AR parameters and the exogenous parameters separately. 

In terms of parameter recovery for the AR coefficients, the ARR2 prior with Minnesota decomposition scales the best with the dimensionality of the exogenous covariates. 
While all other priors deteriorate with dimensionality, the Minnesota ARR2 is the only prior whose performance remains either approximately stable or slightly improves. 

In terms of $\beta$ recovery, stronger differences between the priors emerge. 
In low correlation settings, the Minnesota prior deteriorates in performance in response to increasing dimensions, compared to the ARR2 and RHS priors. 
Yet, it outperforms all other priors again for the highest correlation setting. 
This behaviour is due to the single shared scale in the $\beta$-prior. 
Because in high dimensions, the data can be relatively less informative than the prior, posterior uncertainty is dominated by the prior. 
This creates a uniformly disperse posterior, even for the true zero coefficients. 
Since the true non-zero coefficients are relatively small, parameter recovery appears good as indicated by the RMSE. 
The opposite tendencies are visible for the RHS which does clearly worse with highly correlated covariates, as is a common finding~\citep{piironen2020projective}. 
The spike-and-slab-like prior shape creates the tendency to leave only a small subset of coefficients active at a time for correlated groups. 
The ARR2 priors on the other hand, strike a middle ground between the RHS and Minnesota with good performance throughout. 
To highlight these points, we plot posterior distributions for relevant coefficients in Supplementary Material Section H.

\begin{figure}[!t]
    \centering\input{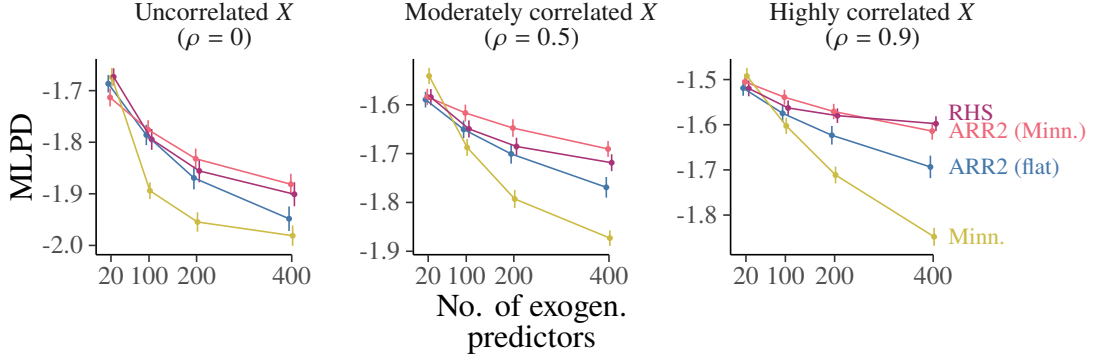}
    \vspace{-\baselineskip}
    \caption{One step ahead leave-future-out mean log predictive density for ARX simulations induced by  our proposed \textcolor{arr2flatcolour}{ARR2 prior with flat concentration} and \textcolor{arr2minncolour}{ARR2 prior with Minnesota concentration}, \textcolor{gaussiancolour}{independent Gaussian priors}, a \textcolor{minnesotacolour}{Minnesota-type prior}, and the \textcolor{rhscolour}{regularised horseshoe prior}. Means and $\pm$ 1 standard error (of 25 repetitions) shown.}
    \label{fig:arx-elpd}
\end{figure}

The results for predictive evaluation are shown in Figure~\ref{fig:arx-elpd}. As with the AR models, the predictive performance generally mirrors the parameter inference quality. The non-locally adaptive shrinkage of the Minnesota prior cause it to perform worst, particularly with increasing dimensions of $X$. Although the RHS tends to concentrate posterior mass on only a few of the relevant covariates, it achieves similar predictive performance to the ARR2 priors which tend to be best in class. All in all, the ARX simulations confirm that the ARR2 prior when including independent covariates, produces competitive predictions while yielding good parameter recovery independent of the dimensionality of the covariate set and their correlation.

\FloatBarrier

\subsection{LTX models} \label{subsec:sim_ltx}
We generate data according to a simple local-linear trend: 
\begin{IEEEeqnarray}{rl} \label{eq:LTX_DGP}
    x_t \;&\sim \normal(0,\Sigma_X) \\
    y_t \;&\sim \normal(x_t^{\prime}\beta + \delta_t,\sigma^2) \\
    \delta_t \;& \sim \normal(\phi \delta_{t-1},\sigma^2_{\delta}), \nonumber
\end{IEEEeqnarray}
where $\Sigma_X$, $\beta$ and $\sigma$ are generated as in Section~\ref{sec:sim_arx}. 
Due to the assumption of independence of states and covariates, we now allow $x_{t,i} \;(i \in \{1,\dots,m\})$ to include $p$ lags. 
Hence, the total dimensionality of $X$ is $T\times m p$.  $\beta_i$ for $i \in \{1,\dots,m\}$ is shared among the $p$ lags of $x_{t,i}$, where we mimic the Minnesota style $\ar$ data generating process by $\tilde{\beta}_{i,j} = \beta_i/j^2$ for $j \in \{1,\dots,p\}$. 
Lags are therefore inversely relevant to their order. 
The state $\ar$ parameter, $\phi$, is set to $0.95$. 
We enforce high state dependence, as would be expected in most time-series applications. 
To test the adaptability of the proposed prior to both high and low signal-to-noise scenarios of the states, we generate from the LTX model with high $(\sigma_{\delta} = 1)$, moderate $(\sigma_{\delta} = 0.5)$ and low $(\sigma_{\delta} = 0.1)$ state scale\footnote{The chosen true state variances relative to the observation noise are taken from simulation studies by \citet{fruhwirth2004efficient}}. This results in true $R^2$ of on average 0.25, 0.7 and 0.9 respectively. Next to the Minnesota and flat decompositions of the ARR2 for state-spaces, we also consider the deterministic decomposition proposed in Section~\ref{sec:arr2_derivation_state_space_models}. That is, the $j$th lag's prior weight for the $i$th covariate is set to $\psi_j1/j^2/(\sum_{s=1}^p(1/s^2))$.
Viewing the states as parameters, the number of unknowns grows proportionally with the number of time-points so that parameter recovery and predictions are evaluated now with increasing time-dimension, while the dimensionality of $x_t$ remains fixed.

Previous studies have found that when the true state scale is low compared to the observation noise, state-space models tend to overfit, especially with many states or when the priors on the state scale have little mass on zero~\citep{huber2021inducing,bitto2019achieving}. 
Having mass on zero is needed in order to identify whether states reduce to a constant over time. 
To allow for sufficient mass on zero while remaining relatively uninformative, we set for the competing prior frameworks, $\sigma_{\delta} \sim \normal(0,3)$. 
Priors for the regression component follow the default recommendations for the RHS and Minnesota prior, respectively.
Due to the significance of $\sigma_{\delta}$, we focus the discussion below only on this parameter, although recovery for other parameters mirror these results (see Supplementary Material Section I).

\begin{figure}[!t]
    \centering\input{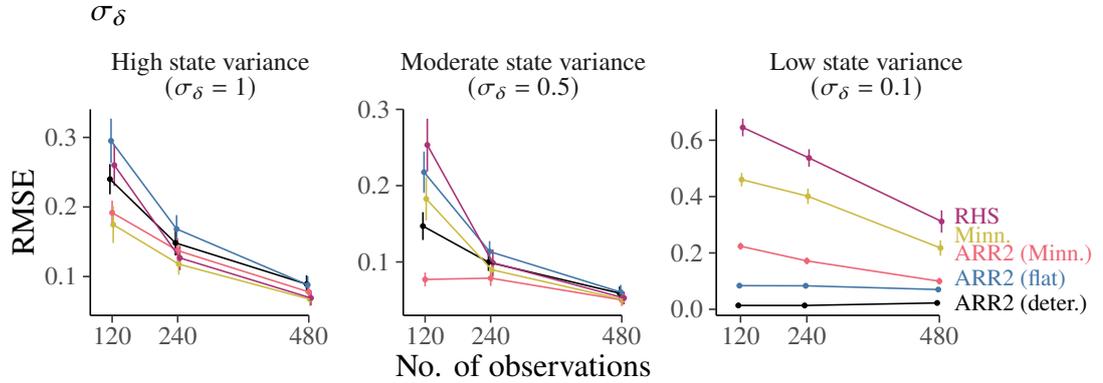}
    \vspace{-\baselineskip}
    \caption{RMSE of posteriors in LTX simulations induced by  our proposed \textcolor{arr2flatcolour}{ARR2 prior with flat concentration} and \textcolor{arr2minncolour}{ARR2 prior with Minnesota concentration}, and ARR2 prior with deterministic decomposition, a \textcolor{minnesotacolour}{Minnesota-type prior}, and the \textcolor{rhscolour}{regularised horseshoe prior}}
    \label{fig:ltx-estim-rmse}
\end{figure}

Figure~\ref{fig:ltx-estim-rmse} plots parameter recovery for $\sigma_{\delta}$. 
When the true state scale is relatively low, the ARR2 priors significantly outperform competing priors. 
The ARR2 deterministic in particular, accurately identifies low $\sigma_{\delta}$. 
As expected, parameter recovery improves with the addition of more observations. However, even with only one latent state variable, it is evident that independently set priors for the states are several orders of magnitude worse compared to any of the ARR2 priors when the sample size and signal from the states are small.
When the true state scale is high, on the other hand, differences between priors decrease. Here the likelihood out-weighs the prior, and the effect of the latter is negligible.

Predictive performance is in line with parameter recovery, showing that the ARR2 priors have on average higher MLPD, particularly in relatively small sample sizes. 
Differences are smaller in predictive space for $\sigma_{\delta}=0.1$ due to the small contribution of the states to predictive variance.

These results clearly show that the joint shrinkage offered by the ARR2 framework is superior both in parameter recovery and predictive performance for state-space models. 
Given that the latent space can be arbitrarily complex, it becomes ever more important to regularise the total variance and decompose according to the contributions to the variance explained. 
And given the large performance benefits of joint shrinkage of state and regression components via $R^2$ in low dimensional state-space models, it is expected that this prior framework may also lead to large performance lifts in high dimensional state-space models such as dynamic regressions \citep{bitto2019achieving}.

\begin{figure}[!t]
    \centering\input{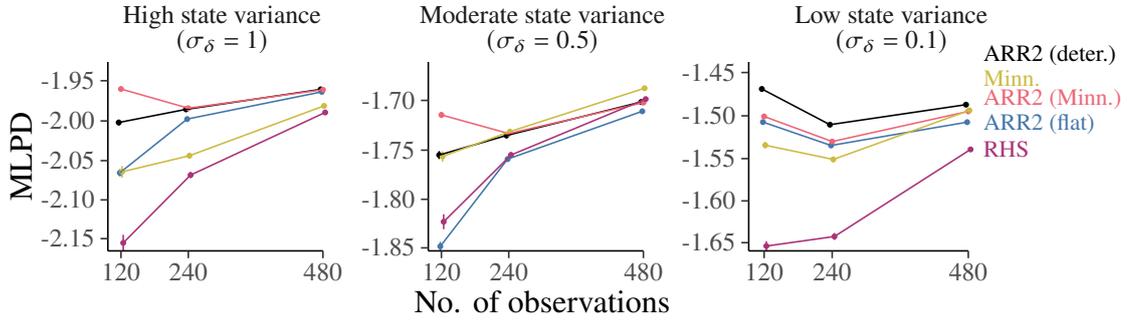}
    \vspace{-\baselineskip}
    \caption{One step ahead leave-future-out mean log predictive density for LTX simulations induced by our proposed \textcolor{arr2flatcolour}{ARR2 prior with flat concentration} and \textcolor{arr2minncolour}{ARR2 prior with Minnesota concentration}, ARR2 prior with deterministic decomposition, a \textcolor{minnesotacolour}{Minnesota-type prior}, and the \textcolor{rhscolour}{regularised horseshoe prior}}
    \label{fig:ltx-lfo}
\end{figure}

\FloatBarrier

\section{Priors in practice: forecasting US CPI inflation}\label{sec:case-study}
For this section, we apply the above methods to forecasting US monthly CPI inflation as a function of monthly macroeconomic covariates which inform on the state of the business cycle and a set of lags. Inflation expectations play a pivotal role in economic decision making such as in central banks where a common task is to accurately predict inflation. A plethora of models have been put forward~\citep{koop_forecasting_2013}, but many studies have found that parsimonious AR, ARX and local trend models produce predictions which are hard to beat, even compared to sophisticated multivariate and non-linear models~\citep{stock_why_2006,athanasopoulos2008varma,hauzenberger2022enhanced}. A commonly cited reason for this is that policy interventions of central banks stabilise movements in inflation toward a targeted range, thus weakening the association between inflation and other macro time-series ~\citep{stock_why_2006}. Economic theory remains inconclusive about structural relationships between time-series of the business cycle and inflation. However, for sound economic analysis, it is still important that any added covariate information, does not lead to a deterioration in predictive performance.

Since LTX models are state-space generalisations of AR and ARX models, we focus in this section on LTX models only.
We define the target as the $1$-month ahead log change in the deseasonalised US CPI index, which we will refer to synonymously as inflation:  
\begin{equation}
    y_{t+1} = \log\left(\frac{\cpi_{t+1}}{\cpi_t}\right).
\end{equation}
For the covariate set, we follow the recent literature by using the FRED-MD dataset~\citep{mccracken2016fred} which gets continuously updated. It maintains a database of 127 monthly time-series that cover price, financial, real economic and survey indicators. All covariates are transformed to stationarity and standardised following the recommendation of \citet{mccracken2016fred}. To capture any lagged effect of the covariates on inflation, we include 12 lags of each covariate making for 1524 covariates $(127\cdot12)$.\footnote{Experiments with only a subset of the \citet{mccracken2016fred} database as made by \citet{hauzenberger2022enhanced} (20 covariates) resulted in nearly identical results.} Predictions are evaluated by MLPD one-step-ahead, based on a rolling window of 240 observations for estimation of model parameters following \citet{hauzenberger2022enhanced}. The initial sample starts January 1981 with the last month forecast being November 2022. We evaluate predictive performance for the ARR2 (Minn.) with deterministic decomposition, ARR2 (Flat) with concentrations $\psi = (1,\dots,1)$, ARR2 (sparse) with concentrations $\psi = (0.1,\dots,0.1)$ , as well as the Minnesota and RHS priors, all defined as in Section~\ref{subsec:sim_ltx}. We firstly compare prior versus posterior tendencies on $R^2$ space before discussing prediction results.

\begin{figure}[!t]
    \centering
    \input{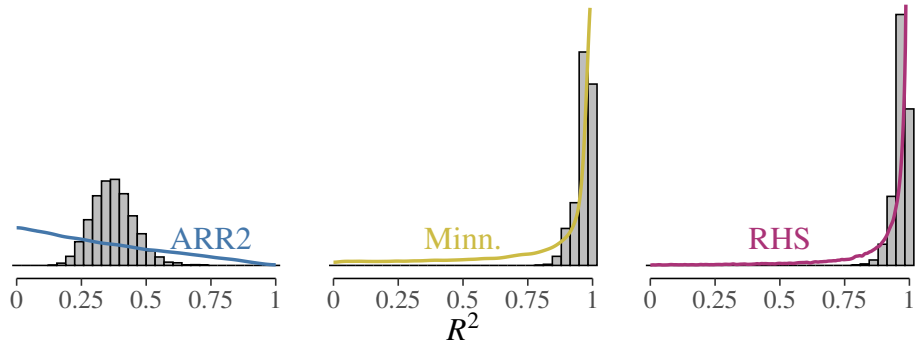}
    \caption{US CPI inflation data. Prior (densities) and posterior (histograms) $R^2$ under different priors. 
    The ARR2 prior is the only prior which regularises the posterior $R^2$ away from one.}
    \label{fig:R2_inflation}
\end{figure}
Figure~\ref{fig:R2_inflation} shows prior predictive and posterior distributions of $R^2$ based on the entire information set (1981-2022) for our set of priors (ARR2 Minnesota shown representatively for the other ARR2 variants). While, by construction, the ARR2 prior, adheres to a $\betadist(1/3,3)$ prior in $R^2$ space, the RHS and Minnesota priors concentrate prior and posterior mass near one. For RHS and Minnesota priors, it is clear that the lack of joint regularisation of the regression and state component does not well regularise $R^2$. Indeed, both models would indicate that nearly 100~$\%$ of the variation in inflation can be explained by the model. 

\begin{figure}[!t]
    \centering
    \input{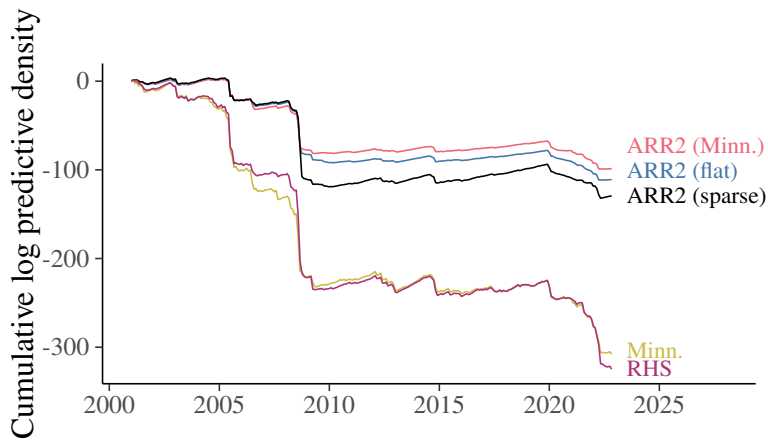}
    \caption{Cumulative log predictive density over time for the US CPI inflation case study. The priors compared are our proposed \textcolor{arr2flatcolour}{ARR2 prior with flat concentration}, \textcolor{arr2minncolour}{ARR2 prior with Minnesota concentration}, and sparse concentration. Also shown are the \textcolor{minnesotacolour}{Minnesota-type prior}, and the \textcolor{rhscolour}{regularised horseshoe prior}.}
    \label{fig:cum_ls_overtime}
\end{figure}

\begin{figure}[!t]
    \centering
    \input{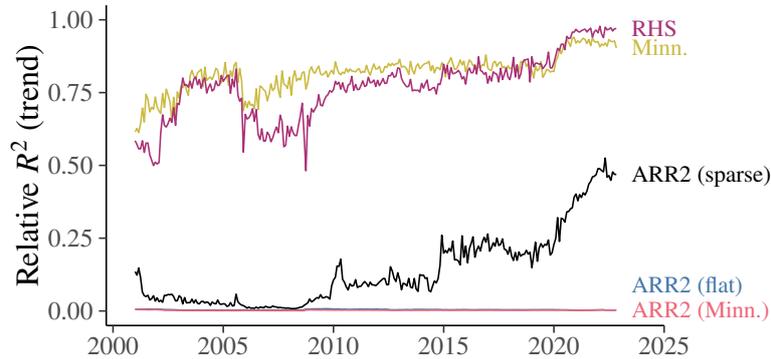}
    \caption{Posterior mean of relative $R^2$ of the trend over time. The priors compared are our proposed \textcolor{arr2flatcolour}{ARR2 prior with flat concentration}, \textcolor{arr2minncolour}{ARR2 prior with Minnesota concentration}, and sparse concentration. Also shown are the \textcolor{minnesotacolour}{Minnesota-type prior}, and the \textcolor{rhscolour}{regularised horseshoe prior}.}
    \label{fig:cum_r2_overtime}
\end{figure}

LFO predictive performance in Figure~\ref{fig:cum_ls_overtime} shows similarly stark differences between the ARR2 priors and the Minnesota/RHS, where the ARR2 priors largely outperform. Here, we show cumulative log predictive density in order to highlight which time-points most severely impact prediction performance. Figure~\ref{fig:cum_ls_overtime} shows that the poor performance of the RHS and Minnesota prior are driven by two significant time-periods: the financial crisis of 2008 and Covid-19 related volatility in inflation where the gap in the lines compared to the ARR2 prior clearly widens. To further unpack this, Figure~\ref{fig:cum_r2_overtime} plots the posterior mean of relative $R^2$ allocated to the state component (total $R^2$ behaves as in Figure~\ref{fig:R2_inflation}) over time. Predictions of the RHS and Minnesota models are driven to a much larger extent by the state component. The latent states capture information of past auto-correlations and thus are not good predictors of changes in inflation which are caused by exogenous shocks to the macroeconomy. The ARR2 priors, in contrast, put more weight on the covariate set leading to more robust predictions.

It is interesting that prior concentrations of the $R^2$ decomposition can lead to quite different posteriors, despite having similar predictions. The ARR2 model which encourages sparsity shows much larger relative $R^2$ allocated to the trend than the flat and Minnesota decompositions. Since smaller concentrations for $\psi$ imply larger variance for any given component, there is likely sufficient variability in the decomposition weights to explore posterior regions closer to the modes found with the RHS and Minnesota priors. 

This tentative application of LTX models to macroeconomic data has largely confirmed the findings established from the simulated experiments: joint regularisation of all model components is crucial to regularise the variance explained by the model, especially of the latent states.  The ARR2 priors lead to superior predictions and estimates of $R^2$ distributions which are more in line with the previous literature~\citep{stock_why_2006}.

\FloatBarrier

\section{Extensions}

The ARR2 prior framework can be extended to other common univariate as well as multivariate time-series models. We present here the $R^2$ definitions of moving-average $\ma(q)$, autoregressive moving-average $\arma(p,q)$, autoregressive distributed lag $\ardl(p,m,q)$ and vector autoregression $\VAR(p)$ models for which we define the corresponding prior hierarchies in Supplementary Material Section G. In the below we present the $R^2$ definition for each model type. Consider all $\epsilon_t$ to be white-noise with variance $\sigma^2$ and that the roots of the characteristic polynomials of any auto-regressive parameters lie outside of the unit circle. Assume initially, as above, that each of these parameters receive an independent normal prior with variance $\lambda^2$.

Define an $\ma(q)$ model as $y_t = \sum_{i=1}^q\varpi_i\epsilon_{t-i} + \epsilon_t$ where $\varpi_i$ are coefficients on the lagged noise terms. Define the linear predictor as $\mu_t = \sum_{i=1}^q \varpi_i \epsilon_{t-i}$. Thus, $\sigma^2_{\mu_t} = \sum_{i=1}^q\lambda_i^2\sigma^2$.  Then, by the same steps in Equations~\ref{eq:marginal-variance-explained}-\ref{eq:marginal-variance-explained-2}, the $R^2$ takes the following form:
\begin{equation}
    R^2_{\mathrm{MA}} = \frac{\sum_{i=1}^q\sigma^2\lambda_i^2}{\sum_{i=1}^q\sigma^2\lambda_i^2 + \sigma^2} = \frac{\sum_{i=1}^q\lambda_i^2}{\sum_{i=1}^q\lambda_i^2 + 1} = \frac{\tau^2}{\tau^2 + 1}.
\end{equation}
Hence, the prior for $\varpi$ need not be scaled by the observation noise in order to cohere with the prior on $R^2$.\footnote{Note that this formulation of the $\ma(q)$ forgoes the complications that arise in equivalently defining the $\ma$~model as an infinite-dimensional $\ar$~model.}

Define an $\arma(p,q)$ model as $y_t = \sum_{i=1}^p \phi_i y_{t-i} + \sum_{j=1}^q\varpi_j\epsilon_{t-j} + \epsilon_t$ where $\mu_t = \sum_{i=1}^p \phi_i y_{t-i} + \sum_{j=1}^q\varpi_j\epsilon_{t-j}$. Then, $\sigma^2_{\mu_t} = \tr{\Lambda\Sigma_{\mu_t}}) = \sum_{i=1}^p \sigma^2_{y_t|\theta} \lambda_i^2  + \sum_{j=p+1}^q\sigma^2\lambda_j^2 $, where $\Sigma_{\mu_t}$ is the variance-covariance matrix of $\mu_t$ conditional on parameters. Then, by the same derivations presented for $\ar$ and $\ma$ models,
\begin{equation}
    R^2_{\mathrm{ARMA}} = \frac{\sum_{i=1}^p \sigma^2_{y_t|\theta} \lambda_i^2  + \sum_{j=p+1}^q\sigma^2\lambda_j^2}{\sum_{i=1}^p \sigma^2_{y_t|\theta} \lambda_i^2  + \sum_{j=p+1}^q\sigma^2\lambda_j^2 + \sigma^2}.
\end{equation}
The variance term of $\mu_t$ conditional on $\theta$ for each lagged white noise term is always equal to $\sigma^2$ and for each lag polynomial term of the target equal to $\sigma^2_{y_t|\theta}$ due to the assumption of stationarity\footnote{One may alternatively specify a large enough $\ar$ or $\ma$ to capture the dynamics of an $\arma$ model.}. The prior for $\varpi$ need again not be scaled by the observation noise in order to cohere with the prior on $R^2$, whereas the prior on $\phi$ does.

Define an $\ardl(p,m,g)$ as $y_t = \sum_{i=1}^p \phi_i y_{t-i} + \sum_{l =1 }^m \sum_{j=1}^g \beta_{l,j}x_{l,t-j} + \epsilon_t$ where $\beta_{l,j}$ are coefficients on the lagged covariate terms which we assume to be in the stationary region. Define $\mu_t = \sum_{i=1}^p \phi_i y_{t-i} + \sum_{l =1 }^m \sum_{j=1}^g \beta_{l,j}x_{l,t-j}$ and by taking the same steps for the derivation of the ARX and ARMA models above:
\begin{equation}
    R^2_{\mathrm{ARDL}} = \frac{\sum_{i = 1}^p \sigma^2_{y_t|\theta}\lambda_i^2 + \sum_{l = 1}^m \sum_{j=1}^g \sigma_{x_l}^2\lambda^2_{l,j} }{\sum_{i = 1}^p \sigma^2_{y_t|\theta}\lambda_i^2 + \sum_{l = 1}^m \sum_{j=1}^g \sigma_{x_l}^2\lambda^2_{l,j} + \sigma^2}.
\end{equation}
For the $\arma$ and $\ardl$ models, the conditional variance terms $\sigma^2_{y_t|\theta}$ may not be analytically available except for the simplest cases \citep{hamilton2020time}. One may again use data-based estimates in-place, as discussed above.

This can be straightforwardly extended to VAR models when the models can be estimated equation-by-equation. Define a $\VAR(p)$ process as vector valued extension to the $\ar(p)$ model, where $y_t \in \mathbb{R}^N$, is generated as $\Phi_0y_t = \Phi_1 y_{t-1} + \cdots + \Phi_{p}y_{t-p} + \epsilon_t$, $\Phi_{i\in\{1,\dotsc,p\}} \in \mathbb{R}^{N\times N}$,  $\epsilon_t \sim \normal(0,\Sigma_y)$,  and $\Phi_0 \in \mathbb{R}^{N \times N}$ is a contemporaneous impact matrix. Using the commonly made (structural) identification restriction of fixing the diagonal to 1 and upper triangular part of $\Phi_0$ to 0, and assuming that \(\Sigma_y\) is diagonal \citep{chan_minnesota-type_2021}, we can estimate the $\VAR(p)$ equation-by-equation. It should be noted that variable ordering in the $\VAR$ may impact predictive performance, so it's vital to check the influence of the ordering for downstream analysis\footnote{A promising avenue may be to consider  order-invariant estimation techniques for $\VAR$s as proposed in \citet{chan2024large}}. The conventional reduced form VAR model, \(y_t = B_1y_{t-1} + \dotsc + B_py_{t-p} + e_{t} \), where \(e_t \sim \normal(0,\tilde{\Sigma})\) can be recovered after structural estimation, by setting for $i$~in~$1,\dotsc,p$, $B_i = \Phi_0^{-1}\Phi_i$ and $\tilde{\Sigma} = \Phi_0^{-1}\Sigma_y\Phi_0^{-1\prime}$. Unlike the structural noise terms, $e_t$ will generally have a full covariance matrix. The equation specific $R^2$ then follows the structure for the $\ardl(p,m,g)$ model which is augmented by contemporaneous variables with $m=N$ and $g=p$. We present further derivations and simulation results for these models in Supplementary Material Section G. We leave further investigation of these important model types to future research.

\section{Discussion}\label{sec:discussion}
In this paper, we propose the ARR2 prior, a new joint prior for time-series models with auto-regressive dynamics motivated from the models' $R^2$ similar to \citet{zhang_bayesian_2022}. We derive this for three sets of popular time-series models: AR, ARX and state-space models with latent auto-regressive dynamics. This prior incorporates prior information of how much variation the model is expected to explain relative to the total variance of the target, and decomposes the total variance by the time-series model components. Compared to popular priors for time-series models, the ARR2 can achieve similar variance decompositions, but unlike other priors, stays constant on $R^2$ space even when adding more covariates or latent states. Decompositions can be informed by the same heuristics popular in time-series analysis (e.g. from economics). As for how to set the prior on $R^2$ space, we suggest using a prior that tends away from 1 to avoid setting large prior mass into the non-stationary domain of the parameter space, and encoding temporal relevance into the concentration hyperparameters.

Simulations show that the ARR2 prior achieves competitive parameter recovery that scales well to higher dimensions. Particularly for state-space models, the joint prior structure  performs excellently in recovering the true state variance. Since this parameter is used to detect significance of the added state components, accurate recoverability is crucial for model building in any Bayesian workflow with latent states. 

An application to US CPI inflation forecasting confirms these findings. The joint regularisation of the proposed priors reduces variability in the state process which results in far superior predictions. Priors which assume independence between the regression and state component result  in unrealistically high $R^2$ and bad predictions during economic crisis periods.

Despite the success of the $R^2$ framework for even relatively simple time-series models, many common time-series models have not been addressed here. For example, a perennial finding in social sciences is that the observation variance varies over time as well. It is not clear at present how to adapt the $R^2$ prior framework to heteroscedastic error covariance models. Additionally, it is also a common finding that signals of latent dynamics or covariate sets vary over time. This motivates extensions of the present framework to time-varying $R^2$, which we are actively investigating. 

Furthermore, the framework presented in this paper is amenable to multi-level and GLMs discussed by \citet{aguilar_intuitive_2023} and \citet{yanchenko_r2d2_2021} which would be easily implementable in popular Bayesian inference software \citep{Strumbelj+etal:2024:software}. 
\subsection*{Acknowledgments}
We acknowledge the computational resources provided by the Aalto Science-IT project, and the support of the Research Council of Finland Flagship programme: Finnish Center for Artificial Intelligence, Research Council of Finland project (340721), and the Finnish Foundation for Technology Promotion. We thank Javier Aguilar and Anna Elisabeth Riha, Galina Potjagailo, Atanas Christev and the participants of the many workshops organised by the Bank of England, University of Helsinki and Heriot-Watt University for helpful comments and discussion.
\bibliography{main}
\onecolumn
\begin{appendices}
\appendix

\section{Induced priors in other spaces}
In this section, we present figures for induced priors on alternative spaces to those presented in the main text to aid understanding of the prior properties.
\begin{figure}[h]
    \centering
\input{tikz/stationarity_plot}    \caption{Induced priors of the largest root of the characteristic polynomial for a AR model with 12 lags. Dotted line at 1 indicates the boundary of the stationary region.}
    \label{fig:prior_charac_root}
\end{figure}

\begin{figure}[h]
    \centering
\input{tikz/induced_priors}    \caption{Induced priors (mean and 90\% interval) of the AR coefficients, Partial auto-correlations and relative \(R^2\) contributions for a model with 12 lags. Note that all of the draws from the independent Gaussian prior implied a non-stationary process, such that the partial autocorrelations were not interpretable and are not shown.}
    \label{fig:induced-priors}
\end{figure}

\FloatBarrier

\section{Alternative $R^2$ priors}

\begin{figure}[h]
    \centering
    % Created by tikzDevice version 0.12.6 on 2024-11-26 14:24:44
% !TEX encoding = UTF-8 Unicode
\begin{tikzpicture}[x=1pt,y=1pt]
\definecolor{fillColor}{RGB}{255,255,255}
\begin{scope}
\definecolor{drawColor}{RGB}{255,255,255}
\definecolor{fillColor}{RGB}{255,255,255}

\path[draw=drawColor,line width= 0.6pt,line join=round,line cap=round,fill=fillColor] (  0.00,  0.00) rectangle (289.08,144.54);
\end{scope}
\begin{scope}
\definecolor{fillColor}{RGB}{255,255,255}

\path[fill=fillColor] (  8.25, 30.69) rectangle (283.58,139.04);
\definecolor{drawColor}{RGB}{86,180,233}

\path[draw=drawColor,line width= 1.1pt,line join=round] ( 20.77, 97.50) --
	( 23.29, 96.87) --
	( 25.82, 96.25) --
	( 28.35, 95.62) --
	( 30.88, 95.00) --
	( 33.41, 94.37) --
	( 35.93, 93.75) --
	( 38.46, 93.12) --
	( 40.99, 92.50) --
	( 43.52, 91.87) --
	( 46.05, 91.25) --
	( 48.58, 90.62) --
	( 51.10, 90.00) --
	( 53.63, 89.37) --
	( 56.16, 88.75) --
	( 58.69, 88.12) --
	( 61.22, 87.50) --
	( 63.75, 86.87) --
	( 66.27, 86.25) --
	( 68.80, 85.62) --
	( 71.33, 85.00) --
	( 73.86, 84.37) --
	( 76.39, 83.75) --
	( 78.92, 83.12) --
	( 81.44, 82.50) --
	( 83.97, 81.87) --
	( 86.50, 81.25) --
	( 89.03, 80.62) --
	( 91.56, 80.00) --
	( 94.09, 79.37) --
	( 96.61, 78.75) --
	( 99.14, 78.12) --
	(101.67, 77.50) --
	(104.20, 76.87) --
	(106.73, 76.24) --
	(109.25, 75.62) --
	(111.78, 74.99) --
	(114.31, 74.37) --
	(116.84, 73.74) --
	(119.37, 73.12) --
	(121.90, 72.49) --
	(124.42, 71.87) --
	(126.95, 71.24) --
	(129.48, 70.62) --
	(132.01, 69.99) --
	(134.54, 69.37) --
	(137.07, 68.74) --
	(139.59, 68.12) --
	(142.12, 67.49) --
	(144.65, 66.87) --
	(147.18, 66.24) --
	(149.71, 65.62) --
	(152.24, 64.99) --
	(154.76, 64.37) --
	(157.29, 63.74) --
	(159.82, 63.12) --
	(162.35, 62.49) --
	(164.88, 61.87) --
	(167.41, 61.24) --
	(169.93, 60.62) --
	(172.46, 59.99) --
	(174.99, 59.37) --
	(177.52, 58.74) --
	(180.05, 58.12) --
	(182.58, 57.49) --
	(185.10, 56.87) --
	(187.63, 56.24) --
	(190.16, 55.62) --
	(192.69, 54.99) --
	(195.22, 54.37) --
	(197.74, 53.74) --
	(200.27, 53.11) --
	(202.80, 52.49) --
	(205.33, 51.86) --
	(207.86, 51.24) --
	(210.39, 50.61) --
	(212.91, 49.99) --
	(215.44, 49.36) --
	(217.97, 48.74) --
	(220.50, 48.11) --
	(223.03, 47.49) --
	(225.56, 46.86) --
	(228.08, 46.24) --
	(230.61, 45.61) --
	(233.14, 44.99) --
	(235.67, 44.36) --
	(238.20, 43.74) --
	(240.73, 43.11) --
	(243.25, 42.49) --
	(245.78, 41.86) --
	(248.31, 41.24) --
	(250.84, 40.61) --
	(253.37, 39.99) --
	(255.90, 39.36) --
	(258.42, 38.74) --
	(260.95, 38.11) --
	(263.48, 37.49) --
	(266.01, 36.86) --
	(268.54, 36.24) --
	(271.06, 35.61);
\definecolor{drawColor}{RGB}{213,94,0}

\path[draw=drawColor,line width= 1.1pt,line join=round] ( 20.77,139.04) --
	( 23.29,134.11) --
	( 25.82,105.62) --
	( 28.35, 93.07) --
	( 30.88, 85.63) --
	( 33.41, 80.59) --
	( 35.93, 76.89) --
	( 38.46, 74.04) --
	( 40.99, 71.75) --
	( 43.52, 69.87) --
	( 46.05, 68.30) --
	( 48.58, 66.95) --
	( 51.10, 65.79) --
	( 53.63, 64.77) --
	( 56.16, 63.88) --
	( 58.69, 63.08) --
	( 61.22, 62.37) --
	( 63.75, 61.73) --
	( 66.27, 61.15) --
	( 68.80, 60.62) --
	( 71.33, 60.14) --
	( 73.86, 59.71) --
	( 76.39, 59.30) --
	( 78.92, 58.93) --
	( 81.44, 58.60) --
	( 83.97, 58.28) --
	( 86.50, 57.99) --
	( 89.03, 57.73) --
	( 91.56, 57.48) --
	( 94.09, 57.25) --
	( 96.61, 57.04) --
	( 99.14, 56.85) --
	(101.67, 56.67) --
	(104.20, 56.51) --
	(106.73, 56.35) --
	(109.25, 56.21) --
	(111.78, 56.09) --
	(114.31, 55.97) --
	(116.84, 55.86) --
	(119.37, 55.77) --
	(121.90, 55.68) --
	(124.42, 55.61) --
	(126.95, 55.54) --
	(129.48, 55.48) --
	(132.01, 55.43) --
	(134.54, 55.39) --
	(137.07, 55.36) --
	(139.59, 55.34) --
	(142.12, 55.32) --
	(144.65, 55.31) --
	(147.18, 55.31) --
	(149.71, 55.32) --
	(152.24, 55.34) --
	(154.76, 55.36) --
	(157.29, 55.39) --
	(159.82, 55.43) --
	(162.35, 55.48) --
	(164.88, 55.54) --
	(167.41, 55.61) --
	(169.93, 55.68) --
	(172.46, 55.77) --
	(174.99, 55.86) --
	(177.52, 55.97) --
	(180.05, 56.09) --
	(182.58, 56.21) --
	(185.10, 56.35) --
	(187.63, 56.51) --
	(190.16, 56.67) --
	(192.69, 56.85) --
	(195.22, 57.04) --
	(197.74, 57.25) --
	(200.27, 57.48) --
	(202.80, 57.73) --
	(205.33, 57.99) --
	(207.86, 58.28) --
	(210.39, 58.60) --
	(212.91, 58.93) --
	(215.44, 59.30) --
	(217.97, 59.71) --
	(220.50, 60.14) --
	(223.03, 60.62) --
	(225.56, 61.15) --
	(228.08, 61.73) --
	(230.61, 62.37) --
	(233.14, 63.08) --
	(235.67, 63.88) --
	(238.20, 64.77) --
	(240.73, 65.79) --
	(243.25, 66.95) --
	(245.78, 68.30) --
	(248.31, 69.87) --
	(250.84, 71.75) --
	(253.37, 74.04) --
	(255.90, 76.89) --
	(258.42, 80.59) --
	(260.95, 85.63) --
	(263.48, 93.07) --
	(266.01,105.62) --
	(268.54,134.11) --
	(271.06,139.04);
\definecolor{drawColor}{gray}{0.60}

\path[draw=drawColor,line width= 1.1pt,line join=round] ( 20.77, 66.56) --
	( 23.29, 66.56) --
	( 25.82, 66.56) --
	( 28.35, 66.56) --
	( 30.88, 66.56) --
	( 33.41, 66.56) --
	( 35.93, 66.56) --
	( 38.46, 66.56) --
	( 40.99, 66.56) --
	( 43.52, 66.56) --
	( 46.05, 66.56) --
	( 48.58, 66.56) --
	( 51.10, 66.56) --
	( 53.63, 66.56) --
	( 56.16, 66.56) --
	( 58.69, 66.56) --
	( 61.22, 66.56) --
	( 63.75, 66.56) --
	( 66.27, 66.56) --
	( 68.80, 66.56) --
	( 71.33, 66.56) --
	( 73.86, 66.56) --
	( 76.39, 66.56) --
	( 78.92, 66.56) --
	( 81.44, 66.56) --
	( 83.97, 66.56) --
	( 86.50, 66.56) --
	( 89.03, 66.56) --
	( 91.56, 66.56) --
	( 94.09, 66.56) --
	( 96.61, 66.56) --
	( 99.14, 66.56) --
	(101.67, 66.56) --
	(104.20, 66.56) --
	(106.73, 66.56) --
	(109.25, 66.56) --
	(111.78, 66.56) --
	(114.31, 66.56) --
	(116.84, 66.56) --
	(119.37, 66.56) --
	(121.90, 66.56) --
	(124.42, 66.56) --
	(126.95, 66.56) --
	(129.48, 66.56) --
	(132.01, 66.56) --
	(134.54, 66.56) --
	(137.07, 66.56) --
	(139.59, 66.56) --
	(142.12, 66.56) --
	(144.65, 66.56) --
	(147.18, 66.56) --
	(149.71, 66.56) --
	(152.24, 66.56) --
	(154.76, 66.56) --
	(157.29, 66.56) --
	(159.82, 66.56) --
	(162.35, 66.56) --
	(164.88, 66.56) --
	(167.41, 66.56) --
	(169.93, 66.56) --
	(172.46, 66.56) --
	(174.99, 66.56) --
	(177.52, 66.56) --
	(180.05, 66.56) --
	(182.58, 66.56) --
	(185.10, 66.56) --
	(187.63, 66.56) --
	(190.16, 66.56) --
	(192.69, 66.56) --
	(195.22, 66.56) --
	(197.74, 66.56) --
	(200.27, 66.56) --
	(202.80, 66.56) --
	(205.33, 66.56) --
	(207.86, 66.56) --
	(210.39, 66.56) --
	(212.91, 66.56) --
	(215.44, 66.56) --
	(217.97, 66.56) --
	(220.50, 66.56) --
	(223.03, 66.56) --
	(225.56, 66.56) --
	(228.08, 66.56) --
	(230.61, 66.56) --
	(233.14, 66.56) --
	(235.67, 66.56) --
	(238.20, 66.56) --
	(240.73, 66.56) --
	(243.25, 66.56) --
	(245.78, 66.56) --
	(248.31, 66.56) --
	(250.84, 66.56) --
	(253.37, 66.56) --
	(255.90, 66.56) --
	(258.42, 66.56) --
	(260.95, 66.56) --
	(263.48, 66.56) --
	(266.01, 66.56) --
	(268.54, 66.56) --
	(271.06, 66.56);
\end{scope}
\begin{scope}
\definecolor{drawColor}{RGB}{213,94,0}

\node[text=drawColor,anchor=base,inner sep=0pt, outer sep=0pt, scale=  0.80] at (152.52, 34.82) {beta(0.5, 1)};
\definecolor{drawColor}{gray}{0.60}

\node[text=drawColor,anchor=base,inner sep=0pt, outer sep=0pt, scale=  0.80] at (152.52, 76.14) {beta(0.5, 2)};
\definecolor{drawColor}{RGB}{86,180,233}

\node[text=drawColor,anchor=base,inner sep=0pt, outer sep=0pt, scale=  0.80] at (102.31,103.69) {beta(0.33, 3)};
\end{scope}
\begin{scope}
\definecolor{drawColor}{RGB}{0,0,0}

\path[draw=drawColor,line width= 0.6pt,line join=round] (  8.25, 30.69) --
	(283.58, 30.69);
\end{scope}
\begin{scope}
\definecolor{drawColor}{gray}{0.20}

\path[draw=drawColor,line width= 0.6pt,line join=round] ( 20.77, 27.94) --
	( 20.77, 30.69);

\path[draw=drawColor,line width= 0.6pt,line join=round] ( 83.34, 27.94) --
	( 83.34, 30.69);

\path[draw=drawColor,line width= 0.6pt,line join=round] (145.91, 27.94) --
	(145.91, 30.69);

\path[draw=drawColor,line width= 0.6pt,line join=round] (208.49, 27.94) --
	(208.49, 30.69);

\path[draw=drawColor,line width= 0.6pt,line join=round] (271.06, 27.94) --
	(271.06, 30.69);
\end{scope}
\begin{scope}
\definecolor{drawColor}{gray}{0.30}

\node[text=drawColor,anchor=base,inner sep=0pt, outer sep=0pt, scale=  0.88] at ( 20.77, 19.68) {0.00};

\node[text=drawColor,anchor=base,inner sep=0pt, outer sep=0pt, scale=  0.88] at ( 83.34, 19.68) {0.25};

\node[text=drawColor,anchor=base,inner sep=0pt, outer sep=0pt, scale=  0.88] at (145.91, 19.68) {0.50};

\node[text=drawColor,anchor=base,inner sep=0pt, outer sep=0pt, scale=  0.88] at (208.49, 19.68) {0.75};

\node[text=drawColor,anchor=base,inner sep=0pt, outer sep=0pt, scale=  0.88] at (271.06, 19.68) {1.00};
\end{scope}
\begin{scope}
\definecolor{drawColor}{RGB}{0,0,0}

\node[text=drawColor,anchor=base,inner sep=0pt, outer sep=0pt, scale=  1.10] at (145.91,  7.64) {$R^2$};
\end{scope}
\end{tikzpicture}
    \caption{Density plots of different recommended prior distributions for $R^2$.}
    \label{fig:r2-alternative}
\end{figure}

\FloatBarrier

\section{Example implementation of the ARR2 prior in Stan}\label{appendix:stan}
We show below how one might implement the ARR2 prior in \proglang{Stan} \citep{carpenter_stan_2017,stan}. A non-centred parameterisation is also provided at: \url{https://github.com/n-kall/arr2}.
    \begin{minted}[xleftmargin=20pt,linenos,escapeinside=||]{stan}
data {
  int<lower=1> T; // number of time points
  vector[T] Y; // observations
  int<lower=0> p; // AR order
  // concentration vector of the Dirichlet prior
  vector<lower=0>[p] cons;
  // data for the R2D2 prior
  real<lower=0> mean_R2; // mean of the R2 prior
  real<lower=0> prec_R2; // precision of the R2 prior
  real<lower=0> sigma_sd; // sd of sigma prior
  // variance estimates of y
  real<lower=0> var_y;
}
parameters {
  vector[p] phi; // AR coefficients
  simplex[p] psi; // decomposition simplex
  real<lower=0, upper=1> R2; // coefficient of determination
  real<lower=0> sigma; // observation model sd
}
transformed parameters {
  real<lower=0> tau2 = R2 / (1 - R2); // Equation 15
  vector[T] mu = rep_vector(0.0, T);
  for (t in (p+1):T) {
    for (i in 1:p) {
      mu[t] += phi[i] * Y[t-i]; // Equation 13
    }
  }
}
model {
  // priors
  phi ~ normal(0, sqrt(sigma^2/var_y * tau2 * psi)); // Equation 14
  R2 ~ beta(mean_R2 * prec_R2, (1 - mean_R2) * prec_R2); // Equation 16
  sigma ~ normal(0, sigma_sd); // Equation 17
  psi ~ dirichlet(cons); // Equation 18
  // likelihood
  Y ~ normal_lpdf(mu, sigma); // Equation 12
}
    \end{minted}

%\newpage

%     \begin{lstlisting}[caption=Stan code for ARR2,language=C, escapechar=|]
% data {
%   int<lower=1> T; // number of time points
%   vector[T] Y; // observations
%   int<lower=0> p; // AR order
%   // concentration vector of the Dirichlet prior
%   vector<lower=0>[p] cons;
%   // data for the R2D2 prior
%   real<lower=0> mean_R2;  // mean of the R2 prior
%   real<lower=0> prec_R2;  // precision of the R2 prior
%   real<lower=0> sigma_sd; // sd of sigma prior
%   // variance estimates of y
%   real<lower=0> var_y;
% }
% parameters {
%   vector[p] phi; // AR coefficients
%   simplex[p] psi; // decomposition simplex
%   real<lower=0, upper=1> R2; // coefficient of determination
%   real<lower=0> sigma; // observation model sd
% }
% transformed parameters {
%   real<lower=0> tau2 = R2 / (1 - R2); // Equation |\ref{eq:arr2_r2_transform}|
%   vector[T] mu = rep_vector(0.0, T);
%   for (t in (p+1):T) {
%     for (i in 1:p) {
%       mu[t] += phi[i] * Y[t-i]; // Equation |\ref{eq:arr2_linear_predictor}|
%     }
%   }
% }
% model {
%   // priors
%   phi ~ normal(0, sqrt(sigma^2/var_y * tau2 * psi)); // Equation |\ref{eq:arr2_phi_prior}|
%   R2 ~ beta(mean_R2 * prec_R2, (1 - mean_R2) * prec_R2); // Equation |\ref{eq:arr2_r2_prior}|
%   sigma ~ normal(0, sigma_sd); // Equation |\ref{eq:arr2_sigma_prior}|
%   psi ~ dirichlet(cons); // Equation |\ref{eq:arr2_dirichlet}|
%   // likelihood
%   Y ~ normal_lpdf(mu, sigma); / Equation |\ref{eq:arr2_likelihood}|
% }
%     \end{lstlisting}
%
\section{Alternative ARX(P,K) priors}\label{appendix:arx-alternatives}
\subsection{Independent Gaussians}
\begin{IEEEeqnarray}{rl}
    y_t \;& \sim\normal (\mu_t,\sigma^2),\quad t=p+1,\dotsc,T \\
    \mu_t \;&= \sum_{i=1}^p\phi_iy_{t-i} + x_t^{\prime}\beta  \\
    \phi_i \;&\sim \normal(0, \sigma_\phi^2), \quad \beta_k \sim \normal(0,\sigma_k^2) \\
    \sigma_\phi^2\;&\sim \pi(\sigma^2),\quad \sigma^2\sim \pi(\sigma^2)
\end{IEEEeqnarray}
\subsection{A Minnesota-style prior}
\begin{IEEEeqnarray}{rl}
    y_t \;& \sim\normal (\mu_t,\sigma^2),\quad t=p+1,\dotsc,T \\
    \mu_t \;&= \sum_{i=1}^p\phi_iy_{t-i} + x_t^{\prime}\beta  \\
    \phi_i \;&\sim \normal(0, \kappa_1 / i^2), \quad \beta_j \sim N(0,\frac{\sigma_y^2}{\sigma_{x_j}^2}\kappa_2),\quad j=1,\dotsc,m \\
    \kappa_1\;&\sim \gammadist(1,1/0.04), \quad \kappa_2\sim \gammadist(1,1/0.04^2), \quad \sigma^2\sim \pi(\sigma^2)
\end{IEEEeqnarray}
Where $\gammadist(1,1/0.04)$ implies the expected prior variance recommended by \citet{carriero_bayesian_2015}. Although hierarchy is fairly standard, this exposition follows \citet{chan_minnesota-type_2021}.
\subsection{The regularised horseshoe prior}
\begin{IEEEeqnarray}{rl}
    y_t \;& \sim\normal (\mu_t,\sigma^2),\quad t=p+1,\dotsc,T \\
    \mu_t \;&= \sum_{i=1}^p\phi_iy_{t-i} + x_t^{\prime}\beta  \\
    \phi_i \;&\sim \normal(0, \tau^2\tilde{\lambda}_j^2), \quad \beta_j \sim \normal(0,\tau^2\tilde{\lambda}^2_j), \\
    i \;& = 1\dotsc,p, j=p+1,\dotsc,p+m \\
    \tilde{\lambda}_j^2 \;&= \frac{c^2\lambda_j^2}{c^2 + \tau^2\lambda_j^2} \\
    \lambda_j \;&\sim \Cauchy_+(0,1) \\
    \tau \;& \sim \Cauchy_+(0,\frac{p_0}{K+p - p_0}\frac{\sigma}{\sqrt{T}}) \\
    \sigma^2\;&\sim \pi(\sigma^2)
\end{IEEEeqnarray}
\section{Derivation of the ARR2 prior for ARX}\label{appendix:arxr2-derivation}
Auto-regressive models with exogenous inputs (ARX) are mathematically constructed as
\begin{equation}
    y_t = \sum_{i=1}^{p} \phi_i y_{t-i} + x_t^{\prime}\beta + \epsilon_t, 
\end{equation}
where $X$ is the data design matrix and $\beta$ the vector of exogenous parameters.

The first point of interest is an understanding of the relationship between model parameters $\phi,x$ and the auto-correlation form of the AR component. As with the usual derivation of the Yule-Walker equations, we being with our process
\begin{equation}
    y_t = \phi_1 y_{t-1} + \cdots + \phi_p y_{t-p} + x_t^{\prime}\beta + \epsilon_t,
\end{equation}
multiply through by $y_{t-h}$,
\begin{equation}
    y_ty_{t-h} = \phi_1 y_{t-1}y_{t-h} + \cdots + \phi_p y_{t-p}y_{t-h} + x_t^{\prime}\beta y_{t-h} + \epsilon_ty_{t-h}
\end{equation}
and take the expectation of all terms conditional on parameters,
\begin{equation}
    \gamma(h) = \phi_1 \gamma(h-1) + \cdots + \phi_p \gamma(h-p) + \E{x_t^{\prime}\beta y_{t-h}} + \E{\epsilon_ty_{t-h}}.
\end{equation}
We are always able to scale the data in the linear model component of the model so that $\E{x_t^{\prime}\beta} = 0$, and thus by linearity for all $h$, $\E{x_t^{\prime}\beta y_{t-h}} = 0$. Further, note that
\begin{equation}
    \E{\epsilon_ty_{t-h}|\theta} = \begin{cases}
        \sigma^2,\quad h=0\\
        0,\quad\text{otherwise}.
    \end{cases}
\end{equation}
As such, we find that the ARX model on standardise exogenous covariates induces the same set of Yule-Walker equations as a pure AR model.

We now return to the definition of $R^2$ as presented by \citet{zhang_bayesian_2022}, namely
\begin{equation}
    R^2 = \frac{\sigma_{\mu_{t}}^2}{\sigma_{\mu_{t}}^2 + \sigma^2}, \label{eq:arx-zhang-r2}
\end{equation}
where we will now look to decompose the total variance into the variance of the location, $\sigma_{\mu_{t}}^2$, and the variance in the observational model, $\sigma^2$. In the case of ARX models, we can further decompose the location into the effects induced by the AR model and the linear model:
\begin{equation}
    \sigma_{y_{t}}^2 = \underbrace{\var{y_{-p}^{\prime}\phi} + \var{x_t^{\prime}\beta}}_{\sigma_{\mu_{t}}^2} +\, \sigma^2.
\end{equation}
Following similar argument to \citet{aguilar_intuitive_2023}, we begin by imposing Gaussian priors over the $\phi$ and $\beta$ parameters. Specifically, we say a priori that
\begin{equation}
    \phi_i \sim \normal(0, \tilde{\lambda}_i^2),\quad \beta_j\sim\normal(0, \tilde{\lambda}_j^2),
\end{equation}
for $i = 1,\dotsc,p$ and $j = p+1,\dotsc,p+m$ are normalised so that
\begin{equation}
    \tilde{\lambda}_i^2 = \frac{\sigma^2}{\sigma_{y_{t}\mid \theta}^2}\lambda_i^2,\quad \tilde{\lambda}_j^2 = \frac{\sigma^2}{\sigma_{x_j}^2}\lambda_j^2.
\end{equation}
By total variance, we can show that
\begin{equation}
    \var{y_{-p}^{\prime}\phi} = \sigma^2\sum_{i=1}^p\lambda_i^2\sigma_{y_{t} \mid \theta}^2. \label{eq:var-arx-ar}
\end{equation}
Similarly, we have
\begin{equation}
    \var{x_t^{\prime}\beta} = \sigma^2\sum_{j=p+1}^{p+m}{\lambda}_i^2\sigma_{x_{j}}^2, \label{eq:var-arx-lin-reg}
\end{equation}
so that by plugging Equations~\ref{eq:var-arx-ar} and~\ref{eq:var-arx-lin-reg} in to Equation~\ref{eq:arx-zhang-r2}, we achieve a formulation of the $R^2$ as 
\begin{equation}
    R^2 = \frac{\sigma^2\sum_{i=1}{\lambda}_i^2 + \sigma^2\sum_{j=p+1}^{p+m}{\lambda}_i^2}{\sigma^2\sum_{i=1}{\lambda}_i^2 + \sigma^2\sum_{j=p+1}^{p+m}{\lambda}_i^2 + \sigma^2}.
\end{equation}
Presently defining
\begin{equation}
    \tau^2 = \sum_{i=1}^{p}{\lambda}_i^2 + \sum_{j=p+1}^{p+m}{\lambda}_i^2,
\end{equation}
we return to our desired form
\begin{equation}
    R^2 =\frac{\tau^2}{\tau^2 + 1},
\end{equation}
from which we can repeat the probabilistic arguments of Section~2 to form a joint predictive prior on ARX models (the so-called ARX-R2 prior). In the formulation provided in-text, we decompose the Dirichlet distribution using $p+1$ concentration parameters, of which $p$ weight on the temporal shrinkage of the AR components, and one remains to allocate variance to the linear regression. Naturally, this can be altered to achieve different decompositions as the user sees fit.
\section{Derivation of the ARR2 prior for state spaces}  \label{appendix:state_space_derivations}
We consider state space models as displayed in Equations 29-30 and maintain Assumptions 1-3. Lean on results from Section~2.4 for the variance of the exogenous linear regression, we turn our attention now to the variance of the state process, 
\begin{IEEEeqnarray}{rl}
    \var{s_t^{\prime}G} \;& = \E[s_t]{\var{s_t^{\prime} G \mid s_t}} + \var{\E[G]{s_t^{\prime}G \mid s_t}}_{s_t} \\
    & = 0 + \var{s_t^{\prime}G}_{s_t} \\
    & = \E[s]{G^{\prime}(s_t-\mu_s)(s_t-\mu_s)^{\prime}G} \\
    & = \tr{\text{cov}(s_t)GG^{\prime}},
\end{IEEEeqnarray}
where $\mu_s$ is $\E{s_t}$. Due to Assumption 3 $\text{cov}(s_t) = \text{cov}(s_{t-1})$. We have from \citet{lutkepohl_new_2005} that $\text{cov}(s_t) = \Phi\text{cov}(s_t) \Phi^{\prime} + \Sigma_s$, which we can solve for using vectorisation: define $A = \Phi \otimes \Phi$, then: 
    \begin{equation} \label{eq:Sigma}
        \text{vec}(\text{cov}(s_t)) = (I_{Q^2 \times Q^2} - A)^{-1}\text{vec}(\Sigma_s),
    \end{equation}
An alternative way to reach the same outcome as in Equation~\ref{eq:Sigma} is via a vector moving average representation (VMA) of the state process. Any stationary $\VAR$ allows a VMA representation, so that a necessary and sufficient condition the condition for Equation~\ref{eq:Sigma} is that $s_t$ is stationary.

To simplify the following derivations, assume, as it is often done in the state space literature \citep{harvey_forecasting_1990}, that $\Phi$ and $\Sigma_s$ are diagonal matrices with entries $(\varphi_1,\cdots,\varphi_Q)$ and $(\sigma^2_{s1},\cdots,\sigma^2_{sQ})$ along the diagonal, respectively. Then, the $i$th diagonal entry of $\text{cov}(s_i)$ is $\frac{\sigma^2_{si}}{1-\varphi_i^2}$. Also we maintain Assumption 2 that restricts the coefficients on the states in the observation equation to be 1. Then the $R^2$ is then written as:
\begin{equation} \label{R2_space}
    R^2 = \frac{\var{\mu_t}}{\var{\mu_t}+\sigma^2} = \frac{\overbrace{ \sigma^2\sum_{j=1}^K\lambda_j^2 + \sigma^2 \sum_{i=1}^q \frac{\sigma^2_{si}}{1-\varphi_i^2} }^{\sigma^2 \tau^2}}{\sigma^2 \tau^2 + \sigma^2}.
\end{equation}
Define the scaled total variance as $\tau^2 = \sum_{j=1}^m\lambda_j^2 + \sum_{i=1}^Q\sigma^2_{si}$. We complete the prior hierarchy as follows:
\begin{equation} \label{eq:state_r2_hier}
    \begin{split}
        \beta_j & \sim \normal(0,\sigma^2\psi_j\tau^2), \quad j = 1,\cdots,m \\
        s_{t,i} & \sim \normal(\varphi_is_{t-1,i},\sigma^2\psi_i (1-\varphi_i^2)\tau^2), \quad i = m+1,\cdots,m+Q \\
        \Phi & \sim \normal(0,\Lambda_S) \\
        \psi & \sim \Dirichlet(\psi) \\
        R^2 & \sim \betadist(\mu_R,\sigma_R), \quad \tau^2 = \frac{R^2}{1-R^2} \\
        \sigma & \sim \pi(\sigma). \\
    \end{split}
\end{equation}
We scale the prior on the states appropriately based on the state transition parameters so as to isolate the state error variances in the computation of $R^2$. This allows us to decompose the $R^2$ entirely based on prior and state variance parameters.
\subsection{Dynamic Regression (DR)}%
\label{appendix:dynamic-regression}
Consider the following dynamic linear regression model

\begin{equation} \label{eq:dr_def}
\begin{split}
    y_t & = x_t^{\prime}\beta_t + \epsilon_t, \quad \epsilon_t \sim \normal(0,\sigma^2) \\
    \beta_{t,j} & = \varphi_j \beta_{t-1,j} + e_{t,j}, \quad e_{t,j} \sim \normal(0,\sigma_{\beta_j}^2), \quad j = 1, \dotsc, m,
\end{split}
\end{equation}
where $\beta_{t,j}$ is a regression coefficient that varies discretely with time according to an $\ar(1)$ process. Define $G = x_t$ from above, then, the marginal $R^2$ is defined as:
\begin{equation}
   R^2 = \frac{\overbrace{ \sigma^2\sum_{j=1}^m\frac{\sigma_{\beta_j}^2}{1-\varphi_j^2}}^{\sigma^2 \tau^2}}{\sigma^2 \tau^2 + \sigma^2} = \frac{\tau^2}{\tau^2 + 1},
\end{equation}
so $\beta_{t,j} - \varphi_j \beta_{t-1,j} \sim \normal\left(0,\sigma^2(1-\varphi_j^2)\psi_j\tau^2\right)$ and $\beta_{0,j} \sim \left(0,\sigma^2(1-\varphi^2_j)\psi_j \tau^2\right)$. Contrary to many previous approaches to Bayesian dynamic regression models, the ARR2 prior makes explicit that the prior variance of the dynamic coefficients is dependent on the observation noise variance and that the total variance of the model is decomposed jointly across all $\beta_t$.

\subsection{Relative $R^2$ of auto-regressive dynamics}%
\label{appendix:relative-r2}
Define the relative $R^2$ as in Section 2.3. Then, conditional on $\theta$, the $i$th relative $R^2$ is
\begin{equation} \label{eq:frequ_rel_R2}
    R^2_{i} \mid \theta = \frac{\var{\phi_{j}y_{t-i}}}{\var{y_t}} = \phi^2_j \frac{\sigma^2/(1-R^2\mid \theta)}{\sigma^2/(1-R^2 \mid \theta)} = \phi^2_i .
\end{equation}
Hence, the relative variance explained by the $i$th lag polynomial is equal to its $i$th-degree partial auto-correlation. The relevance of partial auto-correlations for model building is apparent in frequentist treatment of AR models where they are used to determine appropriate AR orders. 

Equation~\ref{eq:frequ_rel_R2} further motivates a data based estimate for the mean of the $R^2$ prior as well as locations for the Dirichlet components $\psi$. Firstly, notice that a sample based consistent estimator for $\phi$ is $\Gamma^{-1}\rho$ \citep{yule_vii_1927,walker_periodicity_1931} and that $R^2\mid \theta = \sum_{i = 1}^p R^2_i\mid \theta$. Thus, by Equation~\ref{eq:frequ_rel_R2}, a sample based estimator for locations of $R^2$ and $\psi_i$ are $\sum_{i}^p \phi_j^2$ and $\frac{\phi^2_j}{\sum_{i = 1}^p \phi^2_j}$ respectively. In fact for more complicated models such as ARX, $\sum_{i}^p \phi_j^2$ may be used as a crude guess for a soft lower bound on expected $R^2$.

\section{$R^2$ priors for model extensions}
In this section, we outline the $R^2$ definitions as well as the induced prior hierarchies for the moving average ($\ma$), auto-regressive moving-average ($\arma$), auto-regressive distributed-lag ($\ardl$) and the vector-autoregressive ($\VAR$) models. In the model specific $R^2$ definition below, we will assume as in the main text, that any coefficient with index \(i\) has generic prior \(\normal(0,\lambda_i^2)\), all data are stationary and mean zero, and roots of characteristic polynomials of any auto-regressive parameters are outside of the unit circle.
\subsection{$\ma$}
Define the $\ma(q)$ model as
\begin{equation}
    y_t = \sum_{i=1}^q \varpi_i \epsilon_{t-i} + \epsilon_t,
\end{equation}
where all $\epsilon_t$ are independent white noise variables and $\varpi$ is the lagged error coefficient vector. Derivation of the total variance of $\mu_t = \sum_{i=1}^q \varpi_i \epsilon_{t-i}$ is greatly simplified by the white noise assumption that implies $\mathrm{cov}(\epsilon_t,\epsilon_{t-s}) = 0$ for any $s>0$. Hence, by the same steps as presented in the Equations~4-6 in the main paper, and assuming $\varpi_i \sim \normal(0,\lambda_i^2)$ for $i = 1,\dotsc,q$: 
\begin{equation}
    \sigma^2_{\mu_t} = \sum_{i=1}^q \lambda_i^2 \sigma^2  = \sigma^2\tau^2.
\end{equation}
This leads to the $R^2$ definition in Equation~54 of the main paper.
We proceed by specifying the prior below by decomposing the total variance via a Dirichlet prior.
\begin{definition}\label{prop:ma}
    The ARR2 prior for $\ma(q)$ models takes the following form
    \begin{IEEEeqnarray}{rl}
        y_t \;& \sim\normal (\mu_t,\sigma^2),\quad t=q+1,\dotsc,T \\
        \mu_t \;&= \sum_{i=1}^q\varpi_{i}\epsilon_{t-i} \\
        \varpi_i \;&\sim \normal\left(0, \tau^2\psi_{i}\right) \\
        \tau^2 \;&= \frac{R^2}{1 - R^2} \\
        R^2\;&\sim \betadist(\mu_{R^2},\varphi_{R^2}) \\ 
        \sigma^2\;&\sim \pi(\sigma^2) \\
        \psi\;&\sim \Dirichlet(\xi_1,\dotsc,\xi_q).
    \end{IEEEeqnarray}
\end{definition}
\subsection{$\arma$}
Define an $\arma(p,q)$ model as
\begin{equation}
    y_t = \sum_{i=1}^p \phi_i y_{t-i} + \sum_{j=1}^q\varpi_j\epsilon_{t-j} + \epsilon_t,   
\end{equation}
where $p$ notes the lag order of the target and $q$ the lag order of the white noise terms. The conditional variance function is complicated by the appearance of the lagged error terms in the model formulation, where $\mathrm{cov}(y_{t-i},\epsilon_{t-s})\neq0$ for $s\geq i$. Hence, the usual Yule-Walker equations cannot be directly used to solve for the $h$~-the degree autocorrelation function:
\begin{equation}
    \E{y_ty_{t-h}} = \gamma(h) = \phi_1 \gamma(h-1) + \cdots + \phi_p \gamma(h-p) + \E{\sum_{j = 1}^q \epsilon_{t-q}y_{t-h}} + \E{\epsilon_ty_{t-h}},
\end{equation}
where, as above, the expectations are conditional on parameters.
For the derivation of $\sigma^2_{\mu_t} = \var{\sum_{i=1}^p \phi_i y_{t-i} + \sum_{j=1}^q\varpi_j\epsilon_{t-j}}$, we merely need $\gamma(0)$ and $\var{\epsilon_{t-j}|\sigma^2}$. For the former, we refer to the same reasoning as in Section~2.3 to use the marginal data variance of $y_t$, and the latter is always equal to $\sigma^2$ due to the white noise assumption. Hence, $\sigma^2_{\mu_t} = \tr{\Lambda\Sigma_{\mu_t}}) = \sum_{i=1}^p \sigma^2_{y_t|\theta} \lambda_i^2  + \sum_{j=p+1}^q\sigma^2\lambda_j^2$. 

\begin{definition}\label{prop:arma}
    The ARR2 prior for $\arma(p,q)$ models takes the following form
    \begin{IEEEeqnarray}{rl}
        y_t \;& \sim\normal (\mu_t,\sigma^2),\quad t=\max(p,q)+1,\dotsc,T \\
        \mu_t \;&= \sum_{i=1}^p \phi_i y_{t-i} + \sum_{j=1}^q\varpi_j\epsilon_{t-j} \\
        \phi_i \;&\sim \normal\left(0, \frac{\sigma^2}{\sigma_{y_{t}\mid \theta}^2}\tau^2\psi_i \right) \\ 
        \varpi_j \;&\sim \normal\left(0, \tau^2\psi_{j}\right) \\
        \tau^2 \;&= \frac{R^2}{1 - R^2} \\
        R^2\;&\sim \betadist(\mu_{R^2},\varphi_{R^2}) \\ 
        \sigma^2\;&\sim \pi(\sigma^2) \\
        \psi\;&\sim \Dirichlet(\underbrace{\xi_1,\dotsc,\xi_p}_{i=1,\dotsc,p}, \underbrace{\xi_{p+1}, \dotsc,\xi_{p+q}}_{j=1,\dotsc,q}).
    \end{IEEEeqnarray}
\end{definition} 
\subsection{$\ardl$}
Define an $\ardl(p,m,g)$ model as
\begin{equation}
    y_t = \sum_{i=1}^p \phi_i y_{t-i} + \sum_{l =1 }^m \sum_{j=1}^g \beta_{l,j}x_{l,t-j} + \epsilon_t,
\end{equation}
where for notational simplicity, we assume here that for all $m$ covariates, we include the same amount of $g$ lag polynomials. This assumption may be relaxed with some changes to the notation. Assume further that all $x$ are weakly stationary and $\mu_t = \sum_{i=1}^p \phi_i y_{t-i} + \sum_{l =1 }^m \sum_{j=1}^g \beta_{l,j}x_{l,t-j}$. By taking the same steps for the derivation of the ARX and ARMA models above: 
\begin{equation}
    \sigma^2_{\mu_t} =  \sum_{i = 1}^p \sigma^2_{y_t\vert\theta}\lambda_i^2 + \sum_{l = 1}^m \sum_{j=1}^g \sigma_{x_l}^2\lambda^2_{l,j}.
\end{equation}
As for the case of the $\arma$ model, the conditional variance expression $\sigma^2_{y_t|\theta}$ may not have a closed form solution due to the appearance of the lagged covariate terms, but one may resort again to the marginal data variance estimate. This gives the $R^2$ in Equation~56. The induced $R^2$~prior hierarchy is therefore:
\begin{definition}\label{prop:ardl}
    The ARR2 prior for $\ardl(p,m,g)$ models takes the following form
    \begin{IEEEeqnarray}{rl}
        y_t \;& \sim\normal (\mu_t,\sigma^2),\quad t=\max(p,g)+1,\dotsc,T \\
        \mu_t \;&= \sum_{i=1}^p \phi_i y_{t-i} + \sum_{l =1 }^m \sum_{j=1}^g \beta_{l,j}x_{l,t-j} \\
        \phi_i \;&\sim \normal\left(0, \frac{\sigma^2}{\sigma_{y_{t}\mid \theta}^2}\tau^2\psi_i \right) \\
        \beta_{l,j} \;&\sim \normal\left(0, \frac{\sigma^2}{\sigma_{x_l}^2}\tau^2\psi_{l,j}\right) \\
        \tau^2 \;&= \frac{R^2}{1 - R^2} \\
        R^2\;&\sim \betadist(\mu_{R^2},\varphi_{R^2}) \\ 
        \sigma^2\;&\sim \pi(\sigma^2) \\
        \psi\;&\sim\Dirichlet(\underbrace{\xi_1,\dotsc,\xi_p}_{i=1,\dotsc,p}, \underbrace{\xi_{p+1}, \dotsc,\xi_{p+g}}_{l=1,j=1,\dotsc,g},\dotsc,\underbrace{\xi_{p+(m-1)g+1}, \dotsc,\xi_{p+mg}}_{l=m,j=1,\dotsc,g}).
    \end{IEEEeqnarray}
\end{definition}
Notice that due to the weak stationarity assumption of $x$, the coefficients on the lags of each covariate $l \in \{1,\dotsc,m\}$~need only be scaled by $\sigma^2_{x_l}$.
Similar to the discussion for the LTX model, the dimensionality of the estimation problem in Definition~\ref{prop:ardl} can easily get very large as $m$ or $g$ become large. 
One can reduce the computational complexity by decomposing the prior variance at the covariate group-level: $\beta_{l,j} \sim \normal(0,\tilde{\psi}_{l,j}\tau^2) $, where $\tilde{\psi}_{l,j} = w_j \psi_l $, $\psi \sim \Dirichlet(\xi_1,\dots,\xi_m)$ and $\sum_{j=1}^g w_j = 1$ are some deterministic weights for $j \in \{1,\dots,g\}$ and $l \in \{1,\dots,m\}$. In its simplest form, set $w_j=1/g \forall j$ such that the simplex dimensionality reduces to $m+p$ instead of $mg + p$. 
Inspired by a Minnesota-like decomposition for a group of lags, one may then set $w_j$ to $(1/j^2) / (\sum_{s =1 }^g 1/s^2)$.
\subsection{$\VAR$}
Under the assumption of recursive identification and diagonal error-covariance, following \citet{chan_minnesota-type_2021}, we can write the $n$-th equation of the $\VAR(p)$ defined in Section~5 of the main paper as
\begin{equation}
    y_{n,t} = \tilde{y}_{n,t}^{\prime}\alpha_n + \tilde{x}^{\prime}_t\beta_n + \epsilon_{n,t}, \; \epsilon_{n,t} \sim \normal(0,\sigma^2_n),
\end{equation}
where $\tilde{y}_{n,t} = (-y_{1,t},\dotsc,-y_{n-1,t})$, $\tilde{x}_{t} = (y_{t-1}',\dotsc,y_{t-p}')$, \(\alpha_n  = (\alpha_{n,1},\dotsc,\alpha_{n,n-1})^{\prime}\) and $\beta_n = (\beta_{n,1}^{\prime},\dotsc,\beta_{n,p}^{\prime})^{\prime}$. Note that due to the identification scheme, the target of the $n$-th equation is univariate. Define $\mu_{n,t} = \tilde{y}_{n,t}^{\prime}\alpha_n + \tilde{x}^{\prime}_t\beta_n$, then, identical to the derivations of the $\ardl$~model, the total variance definition for the $n$-th equation of the VAR takes the following form: 
\begin{equation}
    \sigma^2_{\mu_{n,t}} = \sum_{j=1}^{n-1}\sigma^2_{y_{j,t}|\theta_n}\lambda_j^2 + \sum_{i = 1}^p \sum_{g=1}^N \sigma^2_{y_{g,t}|\theta_n}\lambda_{i,g}^2.
\end
{equation}
Although multivariate extensions to the Yule-Walkers exist to for consistent estimates of $\sigma^2_{y_{j,t}|\theta_n}$ and $\sigma^2_{y_{g,t}|\theta_n}$ \citep{heaps_enforcing_2022}, one may again use the marginal data variance estimate in place, following similar logic as for the other models.
The equation specific $R^2$ then follows the structure for the $\ardl(p,m,p)$ model, augmented with contemporaneous variables.
The \(R^2_n\) of the $n$-th equation takes the following form:
\begin{equation}
    R^2_n = \frac{\sum_{j=1}^{n-1}\sigma^2_{y_{j,t}|\theta_n}\lambda_j^2 + \sum_{i = 1}^p \sum_{g=1}^N \sigma^2_{y_{g,t}|\theta_n}\lambda_{i,g}^2}{\sum_{j=1}^{n-1}\sigma^2_{y_{j,t}|\theta_n}\lambda_j^2 + \sum_{i = 1}^p \sum_{g=1}^N \sigma^2_{y_{g,t}|\theta_n}\lambda_{i,g}^2 + \sigma^2_n},
\end{equation}
where $\theta_n$ are all parameters that characterise the conditional variance of $y_{n,t}$.
\begin{definition}\label{prop:var}
    Under the structural identification assumption above, the ARR2 prior for $\VAR(p)$ models takes the following form for the $n$-th equation
    \begin{IEEEeqnarray}{rl}
        y_{n,t} \;& \sim\normal (\mu_{n,t},\sigma^2_n),\quad t=p+1,\dotsc,T \\
        \mu_{n,t} \;&= \tilde{y}_{n,t}^{\prime}\alpha_n + \tilde{x}^{\prime}_t\beta_n \\
        \alpha_{n,j} \;&\sim \normal\left(0, \frac{\sigma^2}{\sigma_{y_{j,t}\mid \theta_n}^2}\tau^2\psi_{n,j} \right),\quad j=1,\dotsc,n-1 \\
        \beta_{n,i,g} \;&\sim \normal\left(0, \frac{\sigma^2}{\sigma^2_{y_{g,t}|\theta_n}}\tau^2\psi_{n,i,g}\right), \quad i=1,\dotsc,p \; \text{and} \; g = 1,\dotsc,N \\
        \tau^2 \;&= \frac{R^2}{1 - R^2} \\
        R^2\;&\sim \betadist(\mu_{R^2},\varphi_{R^2}) \\ 
        \sigma^2\;&\sim \pi(\sigma^2) \\
        \psi_n\;&\sim\Dirichlet(\underbrace{\xi_1,\dotsc,\xi_{n-1}}_{j=1,\dotsc,n-1}, \underbrace{\xi_{n}, \dotsc,\xi_{n+N}}_{i=1,g=1,\dotsc,N},\dotsc,\underbrace{\xi_{n+(p-1)N+1}, \dotsc,\xi_{n+pN}}_{i=p,g=1,\dotsc,N}).
    \end{IEEEeqnarray}
\end{definition}
\section{ARX simulations}
\label{appendix:arx-simulations}
\FloatBarrier

\begin{figure}[h]
    \centering
\includegraphics{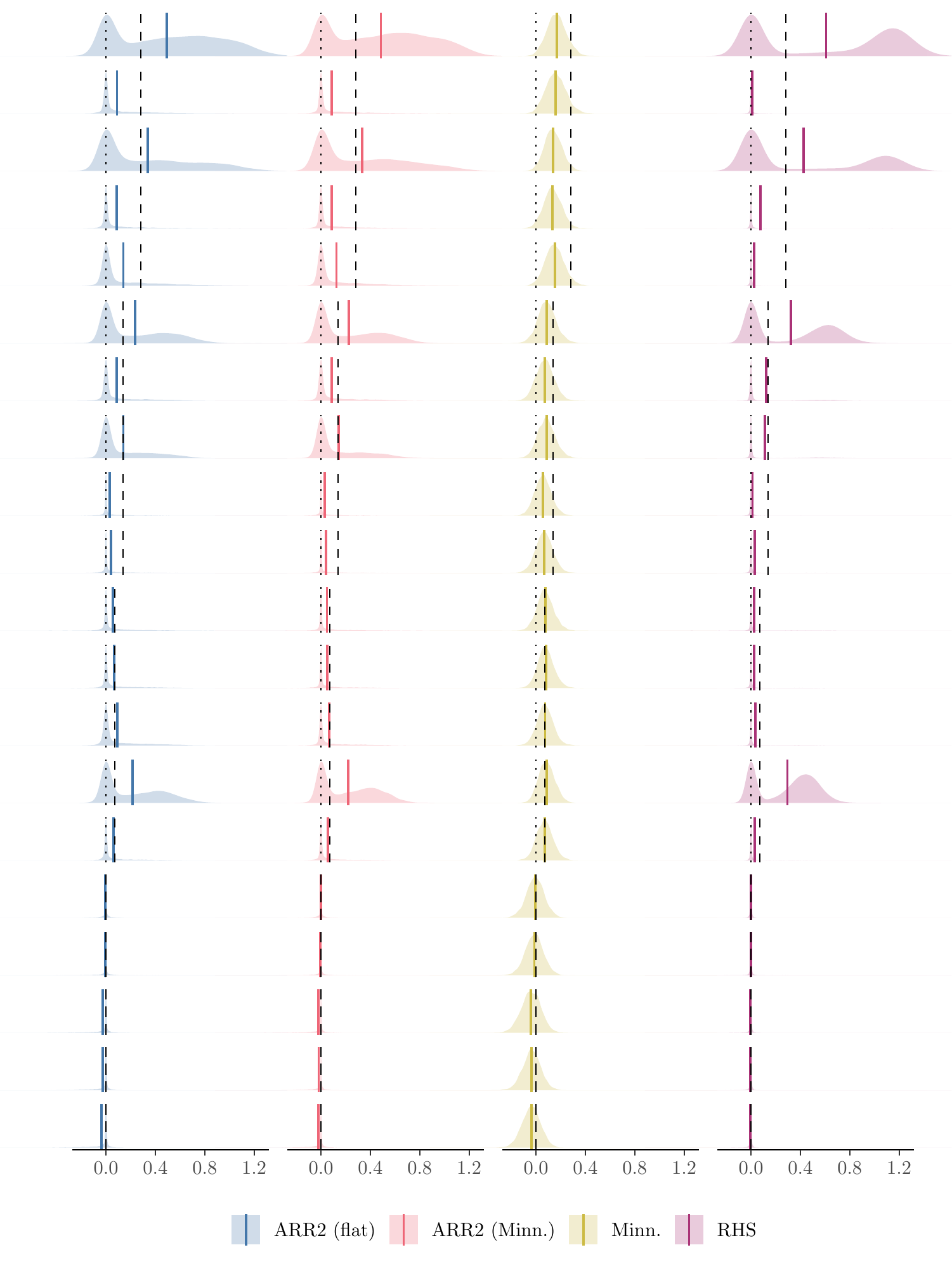}
\caption{Density plots for first 20 \(\beta\) posteriors for \(m = 400, \rho = 0.9\) ARX simulation. The dashed vertical lines indicate the true values; the coloured vertical lines indicate posterior means; the dotted line indicates zero for comparison.}    \label{fig:arx-posterior-density}
\end{figure}

\begin{figure}[h]
    \centering
\includegraphics[width=\textwidth]{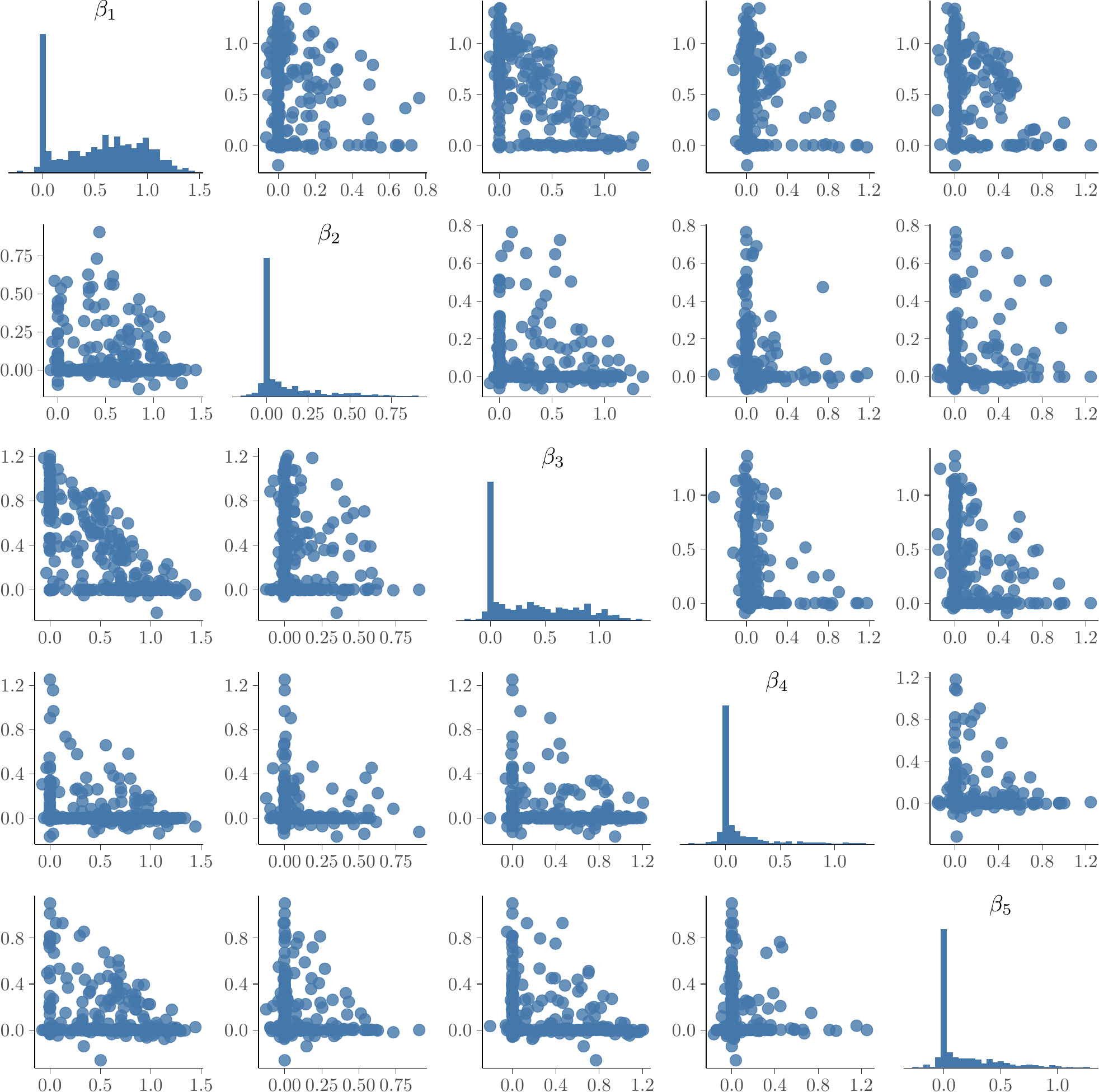}
\caption{Pair plots for first 20 \(\beta\) posteriors for \(m = 400, \rho = 0.9\) ARX simulation. ARR2 (flat) prior.}    \label{fig:arx-pairplot-arr2-flat}
\end{figure}

\begin{figure}[h]
    \centering
\includegraphics[width=\textwidth]{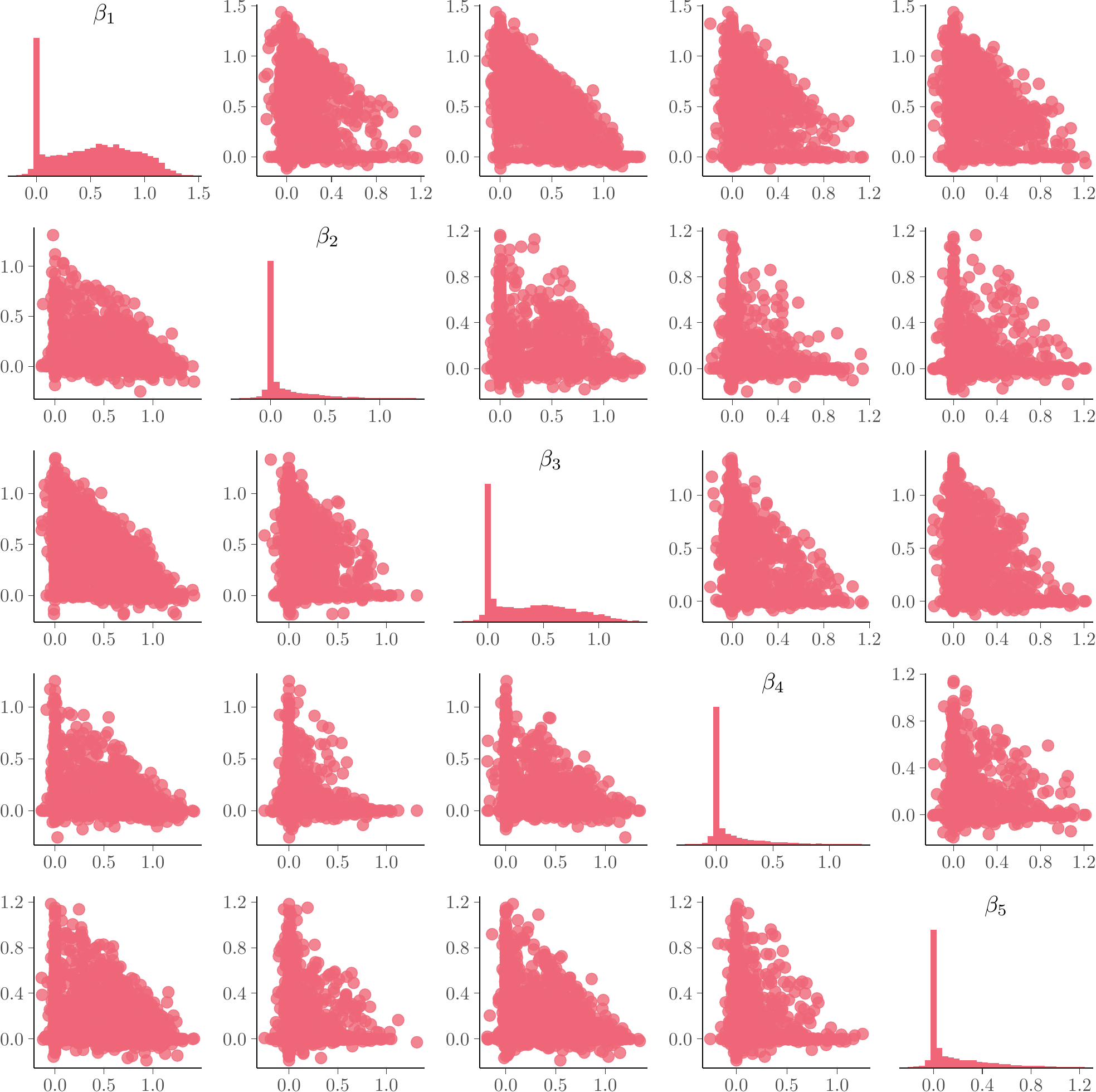}
\caption{Pair plots for first 20 \(\beta\) posteriors for \(m = 400, \rho = 0.9\) ARX simulation. ARR2 (Minnesota) prior.}    \label{fig:arx-pairplot-arr2-minn}
\end{figure}

\begin{figure}[h]
    \centering
\includegraphics[width=\textwidth]{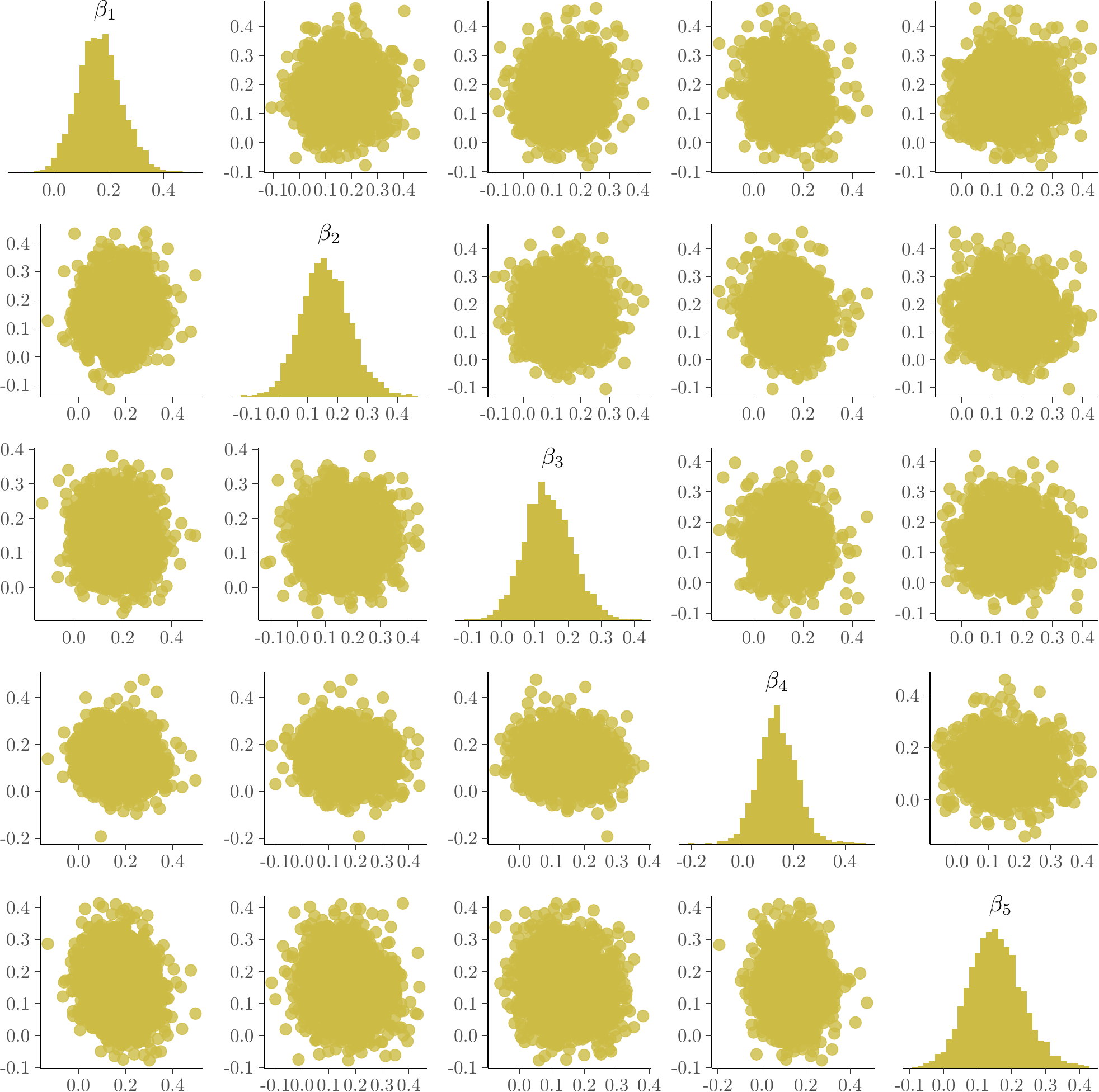}
\caption{Pair plots for first 20 \(\beta\) posteriors for \(m = 400, \rho = 0.9\) ARX simulation. Minnesota prior.}    \label{fig:arx-pairplot-minn}
\end{figure}

\begin{figure}[h]
    \centering
\includegraphics[width=\textwidth]{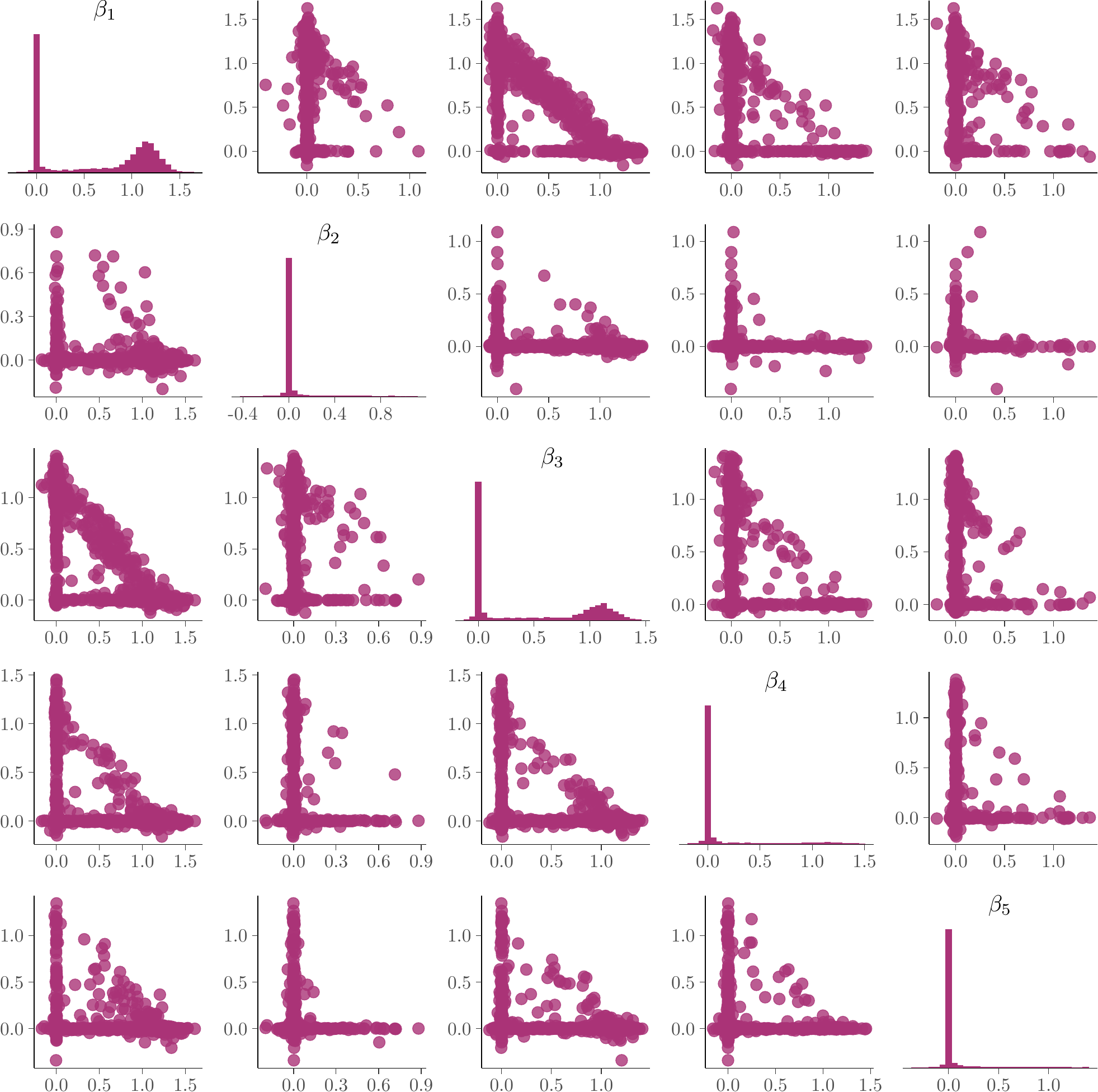}
\caption{Pair plots for first 20 \(\beta\) posteriors for \(m = 400, \rho = 0.9\) ARX simulation. RHS prior.}    \label{fig:arx-pairplot-rhs}
\end{figure}

Figure~\ref{fig:arx-posterior-density} shows the posteriors for one data instance for the ARX simulation. While the Minnesota prior has lower RMSE, the posteriors are wider than the other priors such that the predictions do not perform well. Further, Figures~\ref{fig:arx-pairplot-arr2-flat} to \ref{fig:arx-pairplot-rhs} show the posterior pair plots of the draws for the same models.

\FloatBarrier

\section{LTX simulations}%
\label{appendix:ltx-simulations}

\begin{figure}[h!]
    \centering
\input{tikz/ltx_estim_beta_rmse}
    \caption{RMSE of posteriors induced by  our proposed \textcolor{arr2flatcolour}{ARR2 prior with flat concentration},\textcolor{arr2minncolour}{ARR2 prior with Minnesota concentration}, and deterministic decomposition, a \textcolor{minnesotacolour}{Minnesota-type prior}, and the \textcolor{rhscolour}{regularised horseshoe prior}}
\end{figure}

\begin{figure}[h!]
    \centering
\input{tikz/ltx_estim_tau_rmse}
    \caption{RMSE of posteriors induced by  our proposed \textcolor{arr2flatcolour}{ARR2 prior with flat concentration} and \textcolor{arr2minncolour}{ARR2 prior with Minnesota concentration}, and deterministic decomposition, a \textcolor{minnesotacolour}{Minnesota-type prior}, and the \textcolor{rhscolour}{regularised horseshoe prior}}\end{figure}

\FloatBarrier

\newpage
\section{Forecasting US CPI inflation}

\begin{figure}[h]
    \centering
\input{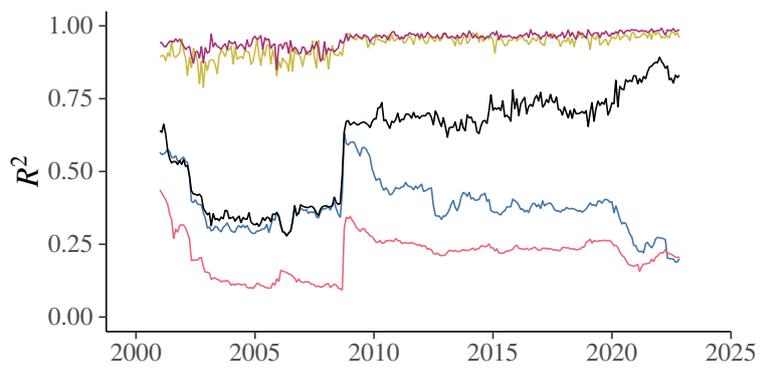}
\caption{Posterior mean of total $R^2$ over time. The priors compared are our proposed \textcolor{arr2flatcolour}{ARR2 prior with flat concentration}, \textcolor{arr2minncolour}{ARR2 prior with Minnesota concentration}, and sparsity-inducing concentration. Also shown are the \textcolor{minnesotacolour}{Minnesota-type prior}, and the \textcolor{rhscolour}{regularised horseshoe prior}.}%
\label{fig:cpi-overall-r2}
\end{figure}

\FloatBarrier

\section{EEG recording during seizure}
\citet{pradoTimeSeriesModeling2021} present a 400 ms EEG recording during a seizure as an example of a dataset which shows quasi-cyclical behaviour (Figure~\ref{fig:eeg-raw}). Here we apply the ARR2 prior to this data and compare the results to the Minnesota and RHS to show that the priors can be used for studying quasi-cyclical behaviour in addition to forecasting. The results of fitting an AR(30) are largely comparable between the priors (Figure~\ref{fig:eeg}), with coefficients after lag 8 being shrunk to zero, or close to zero for all the priors. Similar to the results in Section 4.1, the marginal posteriors of the ARR2 (Minn.) mimic the posterior of the Minnesota prior and the ARR2 (flat), those of the RHS. The \(R^2\), maximum modulus of the characteristic roots and corresponding period mostly align. In line with the findings presented in Figure \ref{fig:prior_charac_root}, we find that the expected maximum modulus for the ARR2 (Minn.) is lower than that of the of the ARR2 (flat). Here the period is defined as \(2 \phi / \operatorname{Arg}(r)\) where \(r\) is the complex root~\citep{pradoTimeSeriesModeling2021}.
\begin{figure}[h]
    \centering
% Created by tikzDevice version 0.12.6 on 2024-11-26 13:48:37
% !TEX encoding = UTF-8 Unicode
\begin{tikzpicture}[x=1pt,y=1pt]
\definecolor{fillColor}{RGB}{255,255,255}
\begin{scope}
\definecolor{drawColor}{RGB}{255,255,255}
\definecolor{fillColor}{RGB}{255,255,255}

\path[draw=drawColor,line width= 0.6pt,line join=round,line cap=round,fill=fillColor] (  0.00,  0.00) rectangle (361.35,130.09);
\end{scope}
\begin{scope}
\definecolor{fillColor}{RGB}{255,255,255}

\path[fill=fillColor] ( 39.04, 30.69) rectangle (355.85,124.59);
\definecolor{drawColor}{RGB}{0,0,0}

\path[draw=drawColor,line width= 0.6pt,line join=round] ( 53.44, 69.07) --
	( 56.35, 66.02) --
	( 59.26, 76.76) --
	( 62.17, 69.36) --
	( 65.08, 85.91) --
	( 67.99,111.03) --
	( 70.90, 93.61) --
	( 73.81, 84.46) --
	( 76.72, 81.70) --
	( 79.63, 76.76) --
	( 82.54, 80.25) --
	( 85.44, 93.02) --
	( 88.35, 67.91) --
	( 91.26, 50.78) --
	( 94.17, 72.70) --
	( 97.08, 83.01) --
	( 99.99, 88.81) --
	(102.90, 70.96) --
	(105.81, 79.81) --
	(108.72, 94.48) --
	(111.63, 96.65) --
	(114.54, 92.88) --
	(117.45, 85.19) --
	(120.35, 61.96) --
	(123.26, 52.96) --
	(126.17, 39.45) --
	(129.08, 63.99) --
	(131.99, 54.55) --
	(134.90,100.72) --
	(137.81,106.96) --
	(140.72,101.44) --
	(143.63,102.90) --
	(146.54, 99.27) --
	(149.45, 83.88) --
	(152.36, 50.63) --
	(155.26, 57.31) --
	(158.17, 67.47) --
	(161.08, 34.95) --
	(163.99, 42.07) --
	(166.90, 64.86) --
	(169.81, 94.04) --
	(172.72,107.40) --
	(175.63, 93.17) --
	(178.54, 86.64) --
	(181.45, 98.11) --
	(184.36,100.14) --
	(187.26, 84.75) --
	(190.17, 90.56) --
	(193.08, 71.25) --
	(195.99, 46.86) --
	(198.90, 51.21) --
	(201.81, 46.57) --
	(204.72, 73.86) --
	(207.63, 84.89) --
	(210.54, 91.43) --
	(213.45, 91.14) --
	(216.36, 96.94) --
	(219.27, 95.93) --
	(222.17,101.74) --
	(225.08,112.91) --
	(227.99, 45.84) --
	(230.90, 60.21) --
	(233.81, 57.89) --
	(236.72, 49.18) --
	(239.63, 59.92) --
	(242.54, 63.99) --
	(245.45, 93.75) --
	(248.36,107.98) --
	(251.27,110.30) --
	(254.18,106.67) --
	(257.08,111.32) --
	(259.99, 47.58) --
	(262.90, 53.25) --
	(265.81, 56.88) --
	(268.72, 67.04) --
	(271.63, 71.39) --
	(274.54, 71.10) --
	(277.45, 70.81) --
	(280.36, 76.04) --
	(283.27,106.24) --
	(286.18,112.62) --
	(289.09,114.80) --
	(291.99, 85.04) --
	(294.90, 68.49) --
	(297.81, 76.62) --
	(300.72, 66.17) --
	(303.63, 49.33) --
	(306.54, 61.38) --
	(309.45, 69.94) --
	(312.36, 72.70) --
	(315.27, 69.65) --
	(318.18, 81.70) --
	(321.09, 93.02) --
	(323.99,120.32) --
	(326.90, 59.49) --
	(329.81, 73.14) --
	(332.72, 77.06) --
	(335.63, 77.20) --
	(338.54, 79.81) --
	(341.45, 67.47);
\end{scope}
\begin{scope}
\definecolor{drawColor}{RGB}{0,0,0}

\path[draw=drawColor,line width= 0.6pt,line join=round] ( 39.04, 30.69) --
	( 39.04,124.59);
\end{scope}
\begin{scope}
\definecolor{drawColor}{gray}{0.30}

\node[text=drawColor,anchor=base east,inner sep=0pt, outer sep=0pt, scale=  0.88] at ( 34.09, 37.18) {-200};

\node[text=drawColor,anchor=base east,inner sep=0pt, outer sep=0pt, scale=  0.88] at ( 34.09, 59.30) {-100};

\node[text=drawColor,anchor=base east,inner sep=0pt, outer sep=0pt, scale=  0.88] at ( 34.09, 81.43) {0};

\node[text=drawColor,anchor=base east,inner sep=0pt, outer sep=0pt, scale=  0.88] at ( 34.09,103.55) {100};
\end{scope}
\begin{scope}
\definecolor{drawColor}{gray}{0.20}

\path[draw=drawColor,line width= 0.6pt,line join=round] ( 36.29, 40.21) --
	( 39.04, 40.21);

\path[draw=drawColor,line width= 0.6pt,line join=round] ( 36.29, 62.33) --
	( 39.04, 62.33);

\path[draw=drawColor,line width= 0.6pt,line join=round] ( 36.29, 84.46) --
	( 39.04, 84.46);

\path[draw=drawColor,line width= 0.6pt,line join=round] ( 36.29,106.58) --
	( 39.04,106.58);
\end{scope}
\begin{scope}
\definecolor{drawColor}{RGB}{0,0,0}

\path[draw=drawColor,line width= 0.6pt,line join=round] ( 39.04, 30.69) --
	(355.85, 30.69);
\end{scope}
\begin{scope}
\definecolor{drawColor}{gray}{0.20}

\path[draw=drawColor,line width= 0.6pt,line join=round] ( 50.53, 27.94) --
	( 50.53, 30.69);

\path[draw=drawColor,line width= 0.6pt,line join=round] (123.26, 27.94) --
	(123.26, 30.69);

\path[draw=drawColor,line width= 0.6pt,line join=round] (195.99, 27.94) --
	(195.99, 30.69);

\path[draw=drawColor,line width= 0.6pt,line join=round] (268.72, 27.94) --
	(268.72, 30.69);

\path[draw=drawColor,line width= 0.6pt,line join=round] (341.45, 27.94) --
	(341.45, 30.69);
\end{scope}
\begin{scope}
\definecolor{drawColor}{gray}{0.30}

\node[text=drawColor,anchor=base,inner sep=0pt, outer sep=0pt, scale=  0.88] at ( 50.53, 19.68) {0};

\node[text=drawColor,anchor=base,inner sep=0pt, outer sep=0pt, scale=  0.88] at (123.26, 19.68) {25};

\node[text=drawColor,anchor=base,inner sep=0pt, outer sep=0pt, scale=  0.88] at (195.99, 19.68) {50};

\node[text=drawColor,anchor=base,inner sep=0pt, outer sep=0pt, scale=  0.88] at (268.72, 19.68) {75};

\node[text=drawColor,anchor=base,inner sep=0pt, outer sep=0pt, scale=  0.88] at (341.45, 19.68) {100};
\end{scope}
\begin{scope}
\definecolor{drawColor}{RGB}{0,0,0}

\node[text=drawColor,anchor=base,inner sep=0pt, outer sep=0pt, scale=  1.10] at (197.45,  7.64) {Time (ms)};
\end{scope}
\begin{scope}
\definecolor{drawColor}{RGB}{0,0,0}

\node[text=drawColor,rotate= 90.00,anchor=base,inner sep=0pt, outer sep=0pt, scale=  1.10] at ( 13.08, 77.64) {EEG signal};
\end{scope}
\end{tikzpicture}
\caption{The first 100 ms of the EEG recording.}.%
\label{fig:eeg-raw}
\end{figure}

\begin{figure}[h]
    \centering
\input{tikz/eeg_plot}
\caption{Posterior AR coefficients, \(R^2\), max modulus and corresponding period for an AR(30) model fit to the EEG data with different priors.}%
\label{fig:eeg}
\end{figure}

\FloatBarrier
\end{appendices}
\end{document}